\pdfoutput=1

\documentclass[iop,revtex4]{emulateapj}
\usepackage{color}
\usepackage{amsmath}
\usepackage{xcolor}
\usepackage{mathrsfs}
\usepackage{natbib}

\shortauthors{Bryan et al.}
\shorttitle{An Excess of Jupiter Analogs in Super-Earth Systems}
\slugcomment{{\sc } Published in AJ}

\begin{document}

\title{An Excess of Jupiter Analogs in Super-Earth Systems}

\author{
Marta L. Bryan\altaffilmark{1},
Heather A. Knutson\altaffilmark{2},
Eve J. Lee\altaffilmark{3},
BJ Fulton\altaffilmark{1,4},
Konstantin Batygin\altaffilmark{2},
Henry Ngo\altaffilmark{5},
Tiffany Meshkat\altaffilmark{1}
}

\altaffiltext{1}{Cahill Center for Astronomy and Astrophysics, California Institute of Technology,
1200 East California Boulevard, MC 249-17, Pasadena, CA 91125, USA}
\altaffiltext{2}{Division of Geological and Planetary Sciences, California Institute of Technology, Pasadena, CA 91125 USA}
\altaffiltext{3}{TAPIR, Walter Burke Institute for Theoretical Physics, Mailcode 350-17, California Institute of Technology, Pasadena, CA 91125, USA}
\altaffiltext{4}{IPAC-NASA Exoplanet Science Institute, Pasadena, CA 91125 USA}
\altaffiltext{5}{National Research Council of Canada, Herzberg Astronomy and Astrophysics, 5071 West Saanich Road, Victoria, BC, V9E 2E7, Canada}

\begin{abstract} 
We use radial velocity observations to search for long-period gas giant companions in systems hosting inner super-Earth ($1-4$ R$_{\Earth}$, $1-10$ M$_{\Earth}$) planets to constrain formation and migration scenarios for this population. We consistently re-fit published RV datasets for 65 stars and find 9 systems with statistically significant trends indicating the presence of an outer companion.  We combine these RV data with AO images to constrain the masses and semi-major axes of these companions.  We quantify our sensitivity to the presence of long-period companions by fitting the sample with a power law distribution and find an occurrence rate of 39$\pm7\%$ for companions $0.5-20$ M$_{\rm Jup}$ and $1-20$ AU.  Half of our systems were discovered by the transit method and half were discovered by the RV method.  While differences in RV baselines and number of data points between the two samples lead to different sensitivities to distant companions, we find that occurrence rates of gas giant companions in each sample are consistent at the 0.5$\sigma$ level.  We compare the frequency of Jupiter analogs in these systems to the equivalent rate from field star surveys and find that Jupiter analogs are more common around stars hosting super-Earths. We conclude that the presence of outer gas giants does not suppress the formation of inner super-Earths, and that these two populations of planets instead appear to be correlated.  We also find that the stellar metallicities of systems with gas giant companions are higher than those without companions, in agreement with the well-established metallicity correlation from RV surveys of field stars.

\end{abstract}

\section{Introduction}
The presence or absence of outer gas giant planets can significantly influence the formation and evolution of planets on interior orbits.  In our own solar system, Jupiter is thought to have played a key role in dynamically re-shaping the outer solar system architecture after the dissipation of the gas disk \citep{2005Natur.435..459T}, driving volatile-rich planetesimals from beyond the ice line onto shorter-period orbits \citep{2006ApJ...643L.131R,2006Icar..184...39O,2012AREPS..40..251M,2017Icar..297..134R}.  At earlier times, the gap in the gas disk created by Jupiter's presence would also have suppressed the flow of solid materials into the inner disk where the terrestrial planets subsequently formed, affecting both the surface density of solids in the inner disk and also the compositions of those solids \citep{2007Icar..191..158M,2012AREPS..40..251M,2014A&A...572A..35L,2016Icar..267..368M,2017arXiv171003809D}. It has even been theorized that an in-and-then-out-again migration by Jupiter and Saturn \citep{2011LPI....42.2585W} might have disrupted planet formation in the inner several AU, therefore explaining why the solar system only hosts relatively small planets between $0.3-2$ AU and none interior to that \citep{2015PNAS..112.4214B}.  

Given the dominant role that gas giant planets played in the early history of the solar system, it is natural to consider their possible influence in exoplanetary systems. Broadly speaking, there are several mechanisms by which outer gas giant planets can influence the formation and evolution of interior planets. Giant planets comparable to or larger than Saturn will open a gap in the gas disk \citep{1986ApJ...309..846L,2006Icar..181..587C,2012ARA&A..50..211K}, potentially suppressing the flow of small solids (``pebbles") to the inner disk.  \citet{2015ApJ...809...94M} find that the rate of planetesimal growth in the inner disk is sensitive to the rate at which pebbles drift radially inward, implying that systems with giant planets should have fewer and less massive planets in the inner region of the disk.  However, the presence of a giant planet will also create local pressure maxima that collect solids, potentially sparking a secondary wave of planet formation \citep{1972fpp..conf..211W,2006ApJ...642..478M,2006MNRAS.373.1619R,2011MNRAS.417.1236H,2012A&A...546A..18M,2016A&A...589A..15S}.

Through resonant transport associated with migration, gas giants can also dynamically excite the population of planetesimals from which rocky planets are forming, increasing the likelihood that collisions will result in disruption rather than accretion \citep{2011LPI....42.2585W,2015PNAS..112.4214B}. However, unless this disrupted material is subsequently accreted onto the host star, this dynamical excitation and disruption of material is not a barrier to rocky planet formation \citep{2017AJ....154..175W}.  Dynamically hot outer gas giants can perturb inner planets onto eccentric and/or inclined orbits, reducing the multiplicity of planets in those systems 
\citep{2017MNRAS.467.1531H,2018MNRAS.tmp.1050P} or leading to orbital instability within a few Myrs in some extreme cases \citep{2017AJ....153..210H}.  These same gas giants can also act as a barrier that prevents smaller planets formed in the outer disk (i.e., beyond the orbit of the gas giants) from migrating inward \citep{2015ApJ...800L..22I}. 

Even if they do not directly influence the formation or dynamical evolution of inner planetary systems, the presence of an outer gas giant planet is in and of itself a statement about the properties of the primordial disk. In the core accretion model \citep{1996Icar..124...62P}, cores must form before the disk gas dissipates in order to acquire massive envelopes.  The well-established correlation between gas giant planet frequency and stellar metallicity for sun-like stars \citep{2005ApJ...622.1102F,2010PASP..122..905J} indicates that the core formation process occurs more readily in metal-rich disks \citep[e.g.][]{2015MNRAS.453.1471D}. The longer lifetime of disks around metal-rich stars also facilitates the formation of both gas giant planets \citep[e.g.][]{2010ApJ...723L.113Y,2010MNRAS.402.2735E} and those at lower masses \citep[e.g.][]{2014Natur.509..593B,2018AJ....155...89P}

Despite the relative richness of theoretical work in this area, we currently have very few observational constraints on the role that outer gas giant planets play in determining the properties of inner planetary systems.  This is largely due to the limited baselines of current surveys: both transit and radial velocity (RV) surveys typically require the observation of one or more complete orbits in order to count a given signal as a secure detection, but even the longest-running surveys have baselines that are shorter than the orbital periods of the solar system gas giants \citep{2008PASP..120..531C,2010Sci...330..653H,2011arXiv1109.2497M,2015ApJ...807...45D,2016ApJ...821...89B,2016ApJ...817..104R,2016ApJ...819...28W}.  Recently, several RV surveys \citep{2016ApJ...819...28W,2016ApJ...817..104R} estimated the frequency of Jupiter analogs, which they defined as 0.3 - 13 M$_{\textnormal{Jup}}$ and 3 - 7 AU \citep{2016ApJ...819...28W} and 0.3 - 3 M$_{\textnormal{Jup}}$ and 3 - 6 AU \citep{2016ApJ...817..104R}, taking into account survey incompleteness at larger separations and smaller masses.  Both surveys found the frequency of Jupiter analogs to be small; \citet{2016ApJ...819...28W} found an occurrence rate of 6.2$^{+2.8}_{-1.6}\%$, while \citet{2016ApJ...817..104R} found an occurrence rate of $\sim$3$\%$.  However, neither of these surveys extended as far as Saturn's orbit, and relatively few of the stars in these two samples have known inner planets.  Of the super-Earth systems examined in this study, we find that only three were included in the \citet{2016ApJ...819...28W} sample, while \citet{2016ApJ...817..104R} did not provide an explicit list of the stars included in their survey.

If we are willing to consider planet candidates with partially observed orbits, we can extend the statistical reach of these surveys to larger orbital separations.  This also allows us to consider systems with inner transiting planets, which typically have shorter photometric and radial velocity baselines (on the order of 1-5 years) \citep[e.g.][]{2014ApJS..210...20M,2014ApJ...783L...6W,2015ApJ...800..135D,2017A&A...602A..88A}.  While the Kepler mission is in principle sensitive to transiting gas giant planets in Jupiter-like orbits \citep{2016ApJ...822....2U,2016AJ....152..206F}, the transit probability for these planets is extremely low and a majority of the long period planet candidates reported to date do not have inner transiting companions.  Alternatively, long-term radial velocity monitoring of systems with known inner planets can provide information on the frequency of outer companions regardless of whether or not they transit their host stars \citep[e.g.][]{2014ApJ...781...28M,2014ApJ...785..126K,2016ApJ...821...89B}.  Although our knowledge of the masses and orbital periods of these objects are incomplete, we can nonetheless search for correlations between inner planet properties and the presence or absence of an outer companion.  

In previous studies we considered the frequency of outer companions in systems with transiting hot Jupiters \citep{2014ApJ...785..126K} and with inner gas giant planets spanning a range of orbital periods \citep{2016ApJ...821...89B}.  In this study we focus on stars known to host one or more super-Earth planets (defined as $1-4$ R$_{\Earth}$ or $1-10$ M$_{\Earth}$, depending on the detection method) located inside 0.5 AU.  These planets dominate the observed population of planets orbiting nearby stars, with 30-50$\%$ of Sun-like stars hosting one or more super-Earths with orbital periods less than 100 days \citep{2010Sci...330..653H,2013ApJ...766...81F,2013PNAS..11019273P,2018arXiv180209526Z}. We identify published RV data for a sample of 65 systems hosting inner super-Earths and use these data to search for long-period gas giant companions.  In section 2 we describe our sample of systems. In section 3 we describe our fits to the RV data, identification of non-planetary sources of RV trends, our calculation of companion probability distributions, and our completeness estimations.  Finally, in section 4 we discuss the occurrence rate of gas giant companions in our sample and implications of our results.

\section{Observations}
We collected published radial velocity (RV) data for systems with at least one confirmed super-Earth, where we define a super-Earth as a planet with either a mass between $1-10$ M$_{\Earth}$ or a radius between $1-4$ R$_{\Earth}$, depending on the detection technique (Table 1).  We exclude systems with fewer than ten data points and baselines shorter than 100 days, leaving us with 65 systems that meet these criteria (Figure 1).  Of that sample, 34 systems host at least one super-Earth discovered using the transit method, and 31 systems host at least one super-Earth discovered using the RV method.  18 of these systems are single-planet systems, while the remaining 47 are multi-planet systems.  45 planets have both measured masses and radii, and thus measured densities.  We provide a summary of the RV data used in this work in Table 1.  We also include best-fit values for the RV acceleration from our orbital solution fitting as described in the following section in Table 1.  We list the complete set of individual RV measurements used in our analysis in Table 2.

Our choice of 1--4 R$_{\Earth}$ and 1--10 M$_{\Earth}$ for inner ``super-Earths'' is  physically motivated and results in a population that is unlikely to be contaminated with Neptune-mass planets. The shape of the planet occurrence rate as a function of orbital period changes drastically for planets smaller vs. larger than 4 R$_{\Earth}$. Whereas the number of smaller planets rises steeply out to $\sim$10 days and plateaus beyond, larger planets grow more numerous with orbital periods out to at least $\sim$300 days \citep{Dong13,Petigura18}. This suggests that these two populations have distinct formation histories and motivates us to choose 4 R$_{\Earth}$ as the upper bound on our sample of super-Earths.\footnote{Although \citet{Fulton17} recently reported evidence for a bimodal distribution in planet radius  with peaks located at 1--1.7 R$_{\Earth}$ and 1.7--4 R$_{\Earth}$, current evidence suggests that this bimodality is not an outcome of divergent formation histories but is instead likely driven by photoevaporative mass loss on a subset of the most highly irradiated planets \citep{Owen13,Owen17}.} For the subset of transiting planets with measured masses, those with radii larger than 4 R$_{\Earth}$ are also typically much more massive than 10 M$_{\Earth}$ \citep{Petigura17}. In order for planets to attain radii greater than 4 R$_{\Earth}$, they must have gas-to-core mass ratio greater than $\sim$10$\%$ \citep{Lopez14}. In a gas-poor but not gas-empty, environment that allows core assembly by giant impact \citep[e.g.,][]{Lee16}, only cores with masses $>\sim$ 10 M$_{\Earth}$ are expected to end up with gas-to-core mass ratios $>\sim$ 10$\%$ \citep[e.g.,][]{Lee15}.

We next check whether our sample is likely to be contaminated by planets whose true masses are comparable to or larger than that of Neptune (17 M$_{\Earth}$).  Looking at the transiting sample first, we find that 26 systems have at least one super-Earth with a mass measurement (note that these are true mass measurements as opposed to $m$sin$i$).  We find that all 26 of these systems have a super-Earth below Neptune mass (17 M$_{\Earth}$).  With regards to the RV mass cutoff of $m$sin$i<$10 M$_{\Earth}$, we quantify what percentage of these planets might have true masses comparable to or larger than that of Neptune. We drew $10^6$ sets of inclinations from a uniform distribution in cos($i$), calculating true masses for each super-Earth in the RV sample.  When there was more than one super-Earth in a system, we selected the lower-mass one.  We found that 5$\%$ of these mass values were above 17 M$_{\Earth}$, whereas 95$\%$ were below Neptune mass.  We therefore conclude that it is unlikely that our sample contains any Neptune-mass planets.

\begin{figure*}
\begin{tabular}{c c}
\includegraphics[width=0.45\textwidth]{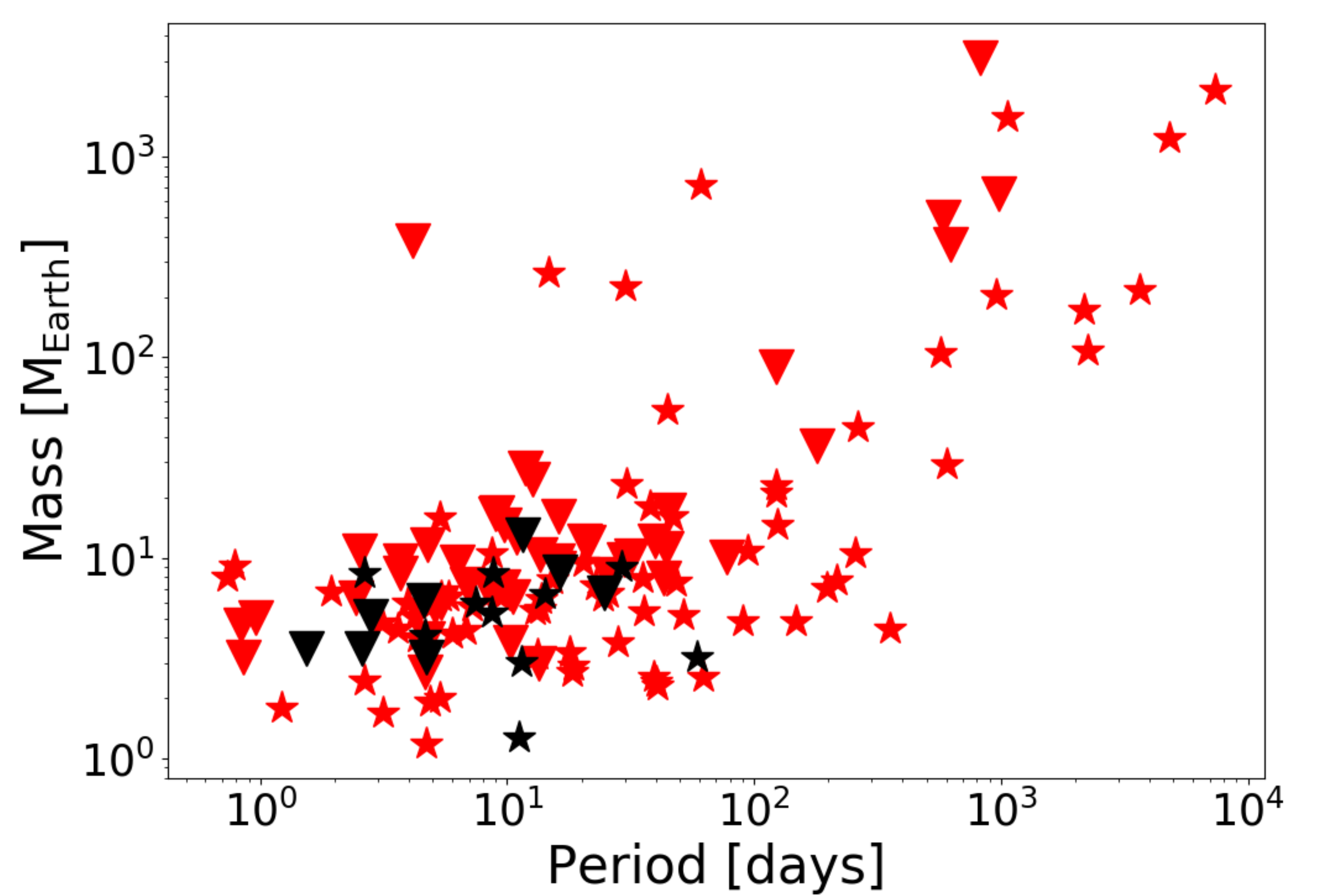}&
\includegraphics[width=0.45\textwidth]{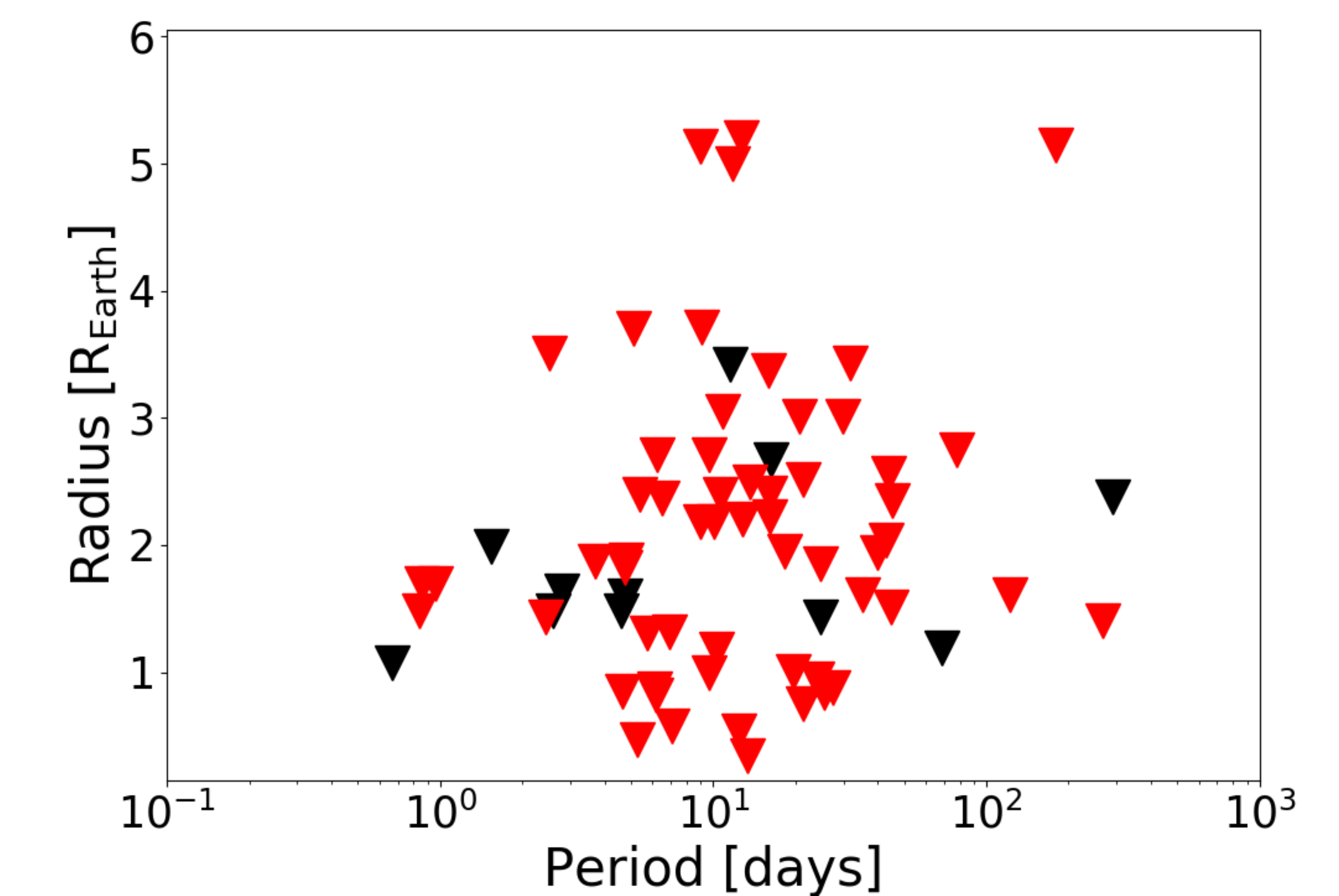}
\end{tabular}
\caption{Confirmed planets on fully resolved orbits from our sample of 65 super-Earth hosting systems.   Each system contains at least one super-Earth ($1-4$ R$_{\Earth}$, $1-10$ M$_{\Earth}$), but some also host well-characterized outer planets with larger masses and radii.  We show the planets with measured masses as a function of period on the left, and planets with measured radii on the right. Systems discovered using the transit method are shown as filled triangles, while systems discovered by the RV method are shown as filled stars.  Multi-planet systems are plotted in red, while single planets are plotted in black.}
\end{figure*}

\LongTables
\begin{deluxetable*}{lcccccccc}
\tabletypesize{\scriptsize}
\tablecaption{Sample of systems}
\tablewidth{0pt}
\tablehead{
\colhead{Target\tablenotemark{3}} & \colhead{M$_{\star}$ (M$_{\odot}$)} & \colhead{[Fe/H]\tablenotemark{1}} & \colhead{N$_{\rm pl}$} & \colhead{Disc. Method} & \colhead{N$_{obs}$} & \colhead{Baseline (days)} & \colhead{Trend (m s$^{-1}$ yr$^{-1}$)}  & \colhead{Ref.}
}
\startdata
Corot-7 & 0.93$\pm$0.03 & 0.09$\pm$0.01 & 2 & Transit & 109 & 357 & 10.95 $\pm$ 7.30 & 9,58 \\
$\textbf{Corot-24}$ & 0.91$\pm$0.09 & 0.30$\pm$0.15 & 2 & Transit & 50 & 1154 & $\textbf{-10.95 $\pm$ 2.92}$ & 10 \\
$\textbf{HD 3167}$ & 0.86$\pm$0.03 & 0.04$\pm$0.05 & 3  & Transit & 251 & 152 & $\textbf{9.96$^{+1.97}_{-2.04}$}$& 37 \\
K2-3 & 0.61$\pm$0.09 & -0.32$\pm$0.13 & 3 & Transit & 72 & 103 & 10.95$^{+6.94}_{-7.67}$ & 12 \\
K2-32 & 0.86$\pm$0.03 & -0.02$\pm$0.01 & 3 & Transit & 74 & 441 & 2.41$^{+2.04}_{-2.01}$ & 14,59 \\
Kepler-10 & 0.91$\pm$0.02 & -0.11$\pm$0.04 & 2 & Transit & 148 & 510 & 3.72$^{+2.04}_{-1.97}$& 8,48  \\
Kepler-20 & 0.91$\pm$0.03 & 0.11$\pm$0.04 & 6 & Transit & 134 & 2262 & -2.01$\pm$1.06 & 13,48, 57 \\
Kepler-21 & 1.41$^{+0.02}_{-0.03}$ & -0.04$\pm$0.04 & 1 & Transit & 122 & 1756 & 0.73 $\pm$ 1.05 & 3,4,48 \\
Kepler-22 & 0.97$\pm$0.06 & -0.20$\pm$0.04 & 1 & Transit & 16 & 373 & 0.84$^{+3.13}_{-3.32}$ & 5,48 \\
Kepler-25 & 1.19$\pm$0.06 & -0.05$\pm$0.04 & 3 & Transit & 62 & 828 & 2.23$^{+2.41}_{-2.30}$ & 2,48 \\
Kepler-37 & 0.80$\pm$0.07 & -0.25$\pm$0.04 & 3 & Transit & 33 & 862 & 0.26 $\pm$ 1.06& 2,48 \\
Kepler-48 & 0.88$\pm$0.06 & 0.26$\pm$0.04 & 4 & Transit & 28 & 1135 & 2.01$^{+3.10}_{-3.32}$ & 2,48 \\
Kepler-62 & 0.69$\pm$0.02 & -0.34$\pm$0.04 & 5 & Transit & 13 & 128 & 60.2$^{+42.0}_{-32.0}$ & 15,48 \\
Kepler-68 & 1.08$\pm$0.05 & 0.14$\pm$.04 & 3 & Transit & 64 & 1207 & 1.68$^{+0.77}_{0.803}$ & 2,48 \\
$\textbf{Kepler-93}$ & 0.91$\pm$0.03 & -0.16$\pm$0.02 & 1 & Transit & 118 & 1892 & $\textbf{12.01$\pm$0.44}$   & 1,2,58 \\
Kepler-94 & 0.81$\pm$0.06 & 0.32$\pm$0.04 & 2 & Transit & 29 & 799 & 28.11$^{+18.62}_{-20.44}$& 2,48 \\
Kepler-95 & 1.08$\pm$0.08 & 0.24$\pm$0.02 & 1 & Transit & 31 & 1078 & 0.62$^{+1.17}_{-1.13}$ & 2,58 \\
Kepler-96 & 1.00$\pm$0.06 & 0.07$\pm$0.02 & 1 & Transit & 26 & 772 & -1.50$^{+1.17}_{-1.10}$  & 2,58 \\
$\textbf{Kepler-97}$ & 0.94$\pm$0.06 & -0.21$\pm$0.02 & 1 & Transit & 20 & 789 & $\textbf{-4.49$^{+1.31}_{-1.35}$}$  & 2,58 \\
Kepler-98 & 0.99$\pm$0.06 & 0.13$\pm$0.02 & 1 & Transit & 22 & 805 & 2.34$^{+2.15}_{-2.04}$ & 2,58 \\
Kepler-99 & 0.79$\pm$0.06 & 0.27$\pm$0.01 & 1 & Transit & 21 & 792 & -2.96$^{+1.35}_{-1.39}$& 2,48 \\
Kepler-100 & 1.08$\pm$0.06 & 0.10$\pm$0.02 & 3 & Transit & 49 & 1221 & 1.06 $\pm$ 0.80 & 2,58 \\
Kepler-102 & 0.81$\pm$0.06 & 0.13$\pm$0.01 & 5 & Transit & 35 & 897 & 1.06$^{+1.13}_{-1.10}$& 2,58 \\
Kepler-103 & 1.09$\pm$0.07 & 0.13$\pm$0.02 & 2 & Transit & 19 & 736 & 2.70 $\pm$ 1.79& 2,58 \\
Kepler-106 & 1.00$\pm$0.06 & -0.07$\pm$0.02 & 4 & Transit & 25 & 1074 & -0.96 $\pm$ 1.3 & 2,58 \\
Kepler-109 & 1.04$\pm$0.06 & -0.01$\pm$0.02 & 2 & Transit & 15 & 1092 & -2.59$^{+2.48}_{-2.81}$& 2,58 \\
Kepler-113 & 0.75$\pm$0.06 & 0.16$\pm$0.01 & 2 & Transit & 24 & 833 & 0.15 $\pm$ 3.65& 2,58 \\
Kepler-131 & 1.02$\pm$0.06 & 0.15$\pm$0.02 & 2 & Transit & 20 & 742 & 0.073$^{+2.11}_{-2.19}$& 2,58 \\
Kepler-406 & 1.07$\pm$0.06 & 0.23$\pm$0.02 & 2 & Transit & 42 & 801 & 0.73 $\pm$ 1.10& 2,58 \\
$\textbf{Kepler-407}$ & 1.00$\pm$0.06 & 0.35$\pm$0.02 & 1 & Transit & 17 & 750 & $\textbf{-156.59 $\pm$}$ 4.02 & 2,58 \\
Kepler-409 & 0.92$\pm$0.06 & 0.05$\pm$0.01 &  1 & Transit & 25 & 175 & 8.76 $\pm$ 6.21 & 2,58 \\
$\textbf{Kepler-454}$ & 1.03$^{+0.04}_{-0.03}$ & 0.22$\pm$0.02 & 2 & Transit & 102 & 1901 & $\textbf{14.56$^{+0.58}_{-0.62}$}$ & 11,58 \\
LHS 1140 & 0.15$\pm$0.02 & -0.24$\pm$0.10 & 1 & Transit & 144 & 386 & 0.44 $\pm$ 1.68 & 6 \\
WASP-47 & 0.99$\pm$0.05 & 0.36$\pm$0.05 & 4 & Transit & 146 & 2340 & -1.31$^{+1.28}_{-2.26}$ & 45, 46, 47\\

55 Cnc & 0.94$\pm$0.05 & 0.36$\pm$0.01 & 5 & RV & 1126 & 8476 & -0.33$^{+0.16}_{-0.15}$ &  16,17,56,59 \\
61 Vir & 0.95$^{+0.04}_{-0.03}$ & -0.04$\pm$0.01 & 3 & RV & 786 & 7060 & -0.31 $\pm$ 0.14 & 18, 19,56,59 \\
GJ 15A & 0.38$\pm$0.06 & -0.32$\pm$0.17 & 1 & RV & 349 & 6215 & -0.44$^{+0.077}_{-0.073}$& 19, 20 \\
GJ 163\tablenotemark{4} & 0.40$\pm$0.04 & -0.01$\pm$0.10 & 3 & RV & 153 & 3068 & -0.12 $\pm$0.16 & 41 \\
GJ 176 & 0.50 & 0.15$\pm$0.17 & 1 & RV & 167 & 5836 & -0.27$^{+0.34}_{-0.35}$ & 19,21,53 \\
$\textbf{GJ 273}$ & 0.29 & -0.17$\pm$0.17 & 2 & RV & 354 & 6855 & $\textbf{1.2$^{+0.066}_{-0.062}$}$ & 19, 22, 53 \\
GJ 433 & 0.48 & -0.22$\pm$0.10 & 1 & RV & 100 & 5476 & -0.22$^{+0.22}_{-0.20}$ & 19,23 \\
GJ 536 & 0.52$\pm$0.05 & -0.08$\pm$0.09 & 1 & RV & 228 & 6128 & -0.13 $\pm$ 0.10 & 19, 24\\
GJ 581\tablenotemark{4} & 0.31$\pm$0.02 & -0.10$\pm$0.17 & 3 & RV & 531 & 5139 & 0.43$^{+0.16}_{-0.15}$ & 19,25,53 \\
$\textbf{GJ 667 C}$\tablenotemark{4} & 0.33$\pm$0.02 & -0.59$\pm$0.10 & 5 & RV & 238 & 4847 & $\textbf{1.79 $\pm$ 0.18}$ & 19,26,27,54\\
$\textbf{GJ 676}$ & 0.71$\pm$0.04 & 0.23$\pm$0.10 & 3 & RV & 127 & 3231 & $\textbf{0 $\pm$ 0}$\tablenotemark{2} & 44 \\
GJ 832 & 0.45$\pm$0.05 & -0.30$\pm$0.20 & 2 & RV & 109 & 5569 & 0.14$^{+0.21}_{-0.22}$ & 32 \\
GJ 876\tablenotemark{4} & 0.33$\pm$0.03 & 0.19$\pm$0.17 & 4 & RV & 401 & 6762 & 1.04 $\pm$ 0.38 & 19,39,53,56 \\
GJ 3138 & 0.68 & -0.30$\pm$0.12 & 3 & RV & 199 & 2932 & 0.20$^{+0.12}_{-0.13}$ & 22 \\
GJ 3293 & 0.42 & 0.02$\pm$0.09 & 4 & RV & 205 & 2311 & -0.11 $\pm$ 0.11 & 22 \\
GJ 3323 & 0.16 & -0.06$\pm$0.17 & 2 & RV & 142 & 4333 & 0.12 $\pm$ 0.11 & 22,53 \\
GJ 3341 & 0.47 & -0.09$\pm$0.09 & 1 & RV & 135 & 1456 & 0.27$\pm$ 0.20 & 28 \\
GJ 3634 & 0.45$\pm$0.05 & -0.10$\pm$0.10 & 1 & RV & 54 & 460 & 9.6$^{+0.95}_{-1.0}$ & 43 \\
GJ 3998 & 0.50$\pm$0.05 & -0.16$\pm$0.09 & 2 & RV & 136 & 869 & -0.66$^{+0.55}_{-0.51}$ & 38 \\
HD 1461 & 1.02 & 0.19$\pm$0.01 & 2 & RV & 921 & 6310 & -0.055$^{+0.16}_{-0.15}$ & 19, 35 \\
HD 7924 & 0.83$^{+0.02}_{-0.04}$ & -0.15$\pm$0.01 & 3 & RV & 906 & 4783 & 0.080 $\pm$ 0.051 & 34,59 \\
HD 20794 & 0.70 & -0.40$\pm$0.01 & 4 & RV & 187 & 2610 & -0.044$^{+0.047}_{-0.044}$ & 31 \\
HD 40307\tablenotemark{4} & 0.77 & -0.31$\pm$0.03 & 5 & RV & 226 & 3811 & 0.55 $\pm$ 0.040 & 35 \\
HD 85512 & 0.69 & -0.33$\pm$0.03 & 1 & RV & 185 & 2745 & 0.32 $\pm$ 0.051 & 31 \\
HD 156668\tablenotemark{4} & 0.77$\pm$0.02 & 0.05$\pm$0.06 & 1 & RV & 527 & 4226 & -0.15 $\pm$ 0.12 & 30,52 \\
HD 175607\tablenotemark{4} & 0.74$\pm$0.05 & -0.62$\pm$0.01 & 1 & RV & 110 & 3390 & 0.13$^{+0.11}_{-0.12}$ & 40 \\
HD 181433 & 0.78 & 0.33$\pm$0.13 & 3 & RV & 107 & 1757 & 1.5$^{+0.98}_{-2.9}$ & 33 \\
HD 215497 & 0.87$\pm$0.02 & 0.23$\pm$0.07 & 2 & RV & 99 & 1842 & -0.24 $\pm$ 0.23 & 29 \\
HD 219134 & 0.81$\pm$0.02 & 0.11$\pm$0.04 & 6 & RV & 1033 & 7421 & -0.45 $\pm$ 0.08 & 19,42,56 \\
Proxima Cen & 0.12$\pm$0.02 & 0.22$\pm$0.03 & 1 & RV & 144 & 4325 & -0.13$^{+0.10}_{-0.095}$& 36,51,55 \\
Wolf 1061 & 0.29 & -0.02$\pm$0.17 & 3 & RV & 187 & 4136 & 0.037$^{+0.058}_{-0.055}$ & 22,53

\enddata

\tablenotetext{1}{We note that uncertainties on the metallicity were not published for systems 61 Vir, GJ 433, GJ 667, Proxima Cen, and GJ 3634.  For these systems we adopt metallicity uncertainties of 0.1 dex.}
\tablenotetext{2}{Because the RV acceleration in GJ 676 has curvature, we fit this long period signal with an orbital solution.  Since this partially resolved orbit and a linear trend are degenerate, we fix the linear trend term in this fit to zero, as well as the eccentricity of this outer companion.}
\tablenotetext{3}{Systems in bold have statistically significant long term trends.}
\tablenotetext{4}{For systems GJ 667, GJ 876, GJ 581, and HD 40307 we fit fewer signals than the published number, and for systems HD 156668, HD 175607, and GJ 163 we fit additional signals.  See section 3.1 for details.}
\tablecomments{References:  (1) \citet{2015ApJ...800..135D}, (2) \citet{2014ApJS..210...20M}, (3) \citet{2016AJ....152..204L}, (4) \citet{2012ApJ...746..123H}, (5) \citet{2012ApJ...745..120B}, (6) \citet{2017Natur.544..333D}, (7) \citet{2015Natur.527..204B}, (8) \citet{2014ApJ...789..154D}, (9) \citet{2009A&A...506..303Q}, (10) \citet{2014A&A...567A.112A}, (11) \citet{2016ApJ...816...95G}, (12) \citet{2015A&A...581L...7A}, (13) \citet{2012ApJ...749...15G}, (14) \citet{2017AJ....153..142P}, (15) \citet{2013Sci...340..587B}, (16) \citet{2012ApJ...759...19E}, (17) \citet{2014MNRAS.441..442N}, (18) \citet{2010ApJ...708.1366V}, (19) \citet{2017AJ....153..208B}, (20) \citet{2014ApJ...794...51H}, (21) \citet{2009A&A...493..645F}, (22) \citet{2017A&A...602A..88A}, (23) \citet{2013A&A...553A...8D}, (24) \citet{2017A&A...597A.108S}, (25) \citet{2009A&A...507..487M}, (26) \citet{2013A&A...556A.126A}, (27) \citet{2012ApJ...751L..16A}, (28) \citet{2015A&A...575A.119A}, (29) \citet{2010A&A...512A..48L}, (30) \citet{2016ApJ...821...89B}, (31) \citet{2011A&A...534A..58P}, (32) \citet{2014ApJ...791..114W}, (33) \citet{2009A&A...496..527B}, (34) \citet{2015ApJ...805..175F}, (35) \citet{2016A&A...585A.134D}, (36) \citet{2016Natur.536..437A}, (37) \citet{2017AJ....154..122C}, (38) \citet{2016A&A...593A.117A}, (39) \citet{2010A&A...511A..21C}, (40) \citet{2016A&A...585A.135M}, (41) \citet{2013A&A...556A.110B}, (42) \citet{2017NatAs...1E..56G}, (43) \citet{2011A&A...528A.111B}, (44) \citet{2016A&A...595A..77S}, (45) \citet{2017AJ....153...70S}, (46) \citet{2016A&A...586A..93N}, (47) \citet{2015ApJ...813L...9D}, (48) \citet{2017AJ....154..107P}, (49) \citet{2009A&A...506..287L}, (50) \citet{2005A&A...442..635B}, (51) \citet{2010A&A...519A.105S}, (52) \citet{2011ApJ...726...73H}, (53) \citet{2012ApJ...748...93R}, (54) \citet{2001A&A...373..159C}, (55) \citet{2018AJ....155...24Z}, (56) \citet{2014ApJS..210....5F}, (57) \citet{2016AJ....152..160B}, (58) \citet{Brewer2018}, (59) \citet{Brewer2016}.}

\end{deluxetable*}

\begin{deluxetable}{lccc}
\tabletypesize{\scriptsize}
\tablecaption{Published RVs Used In This Study}
\tablewidth{0pc}
\tablehead{
\colhead{System} &
\colhead{JD}&
\colhead{RV [m s$^{-1}$]}&
\colhead{$\sigma_{RV}$ [m s$^{-1}$]} 
}
\startdata
Corot-7 & 2454527.5 & 31181.8 & 1.5 \\
Corot-7 & 2454530.6 & 31173.2 & 1.5 \\
Corot-7 & 2454550.5 & 31197.2 & 2.1 \\
Corot-7 & 2454775.8 & 31188.1 & 2.2 \\
Corot-7 & 2454776.7 & 31184.8 & 2.8 \\
Corot-7 & 2454778.7 & 31181.7 & 2.0 \\
Corot-7 & 2454779.7 & 31173.4 & 1.6 \\
Corot-7 & 2454780.7 & 31175.0 & 2.2 \\
Corot-7 & 2454789.8 & 31180.5 & 2.6 \\
Corot-7 & 2454790.8 & 31187.8 & 1.6 

\enddata
\tablecomments{The full set of RVs for each of these systems are available as electronic tables online.}

\end{deluxetable}

 \section{Analysis}

\subsection{RV Fitting}

The presence of a distant companion manifests as a long term trend in the RV data when the orbital period of the companion is significantly longer than the RV baseline.  In order to quantify the significance of these long-term trends, we simultaneously fit for the orbits of the known inner planets as well as a linear trend in each dataset using RadVel \citep{2018PASP..130d4504F}.  After identifying the best-fit solution for each data set, we next carry out a Markov Chain Monte Carlo (MCMC) exploration of the parameter space to determine the uncertainties on each model parameter.  For a system with a single known planet, our model has eight free parameters including six orbital parameters (the planet's velocity semi-amplitude, orbital period, eccentricity, argument of periastron, true anomaly, and an RV zero point), a linear velocity trend, and stellar jitter.

We fit using the basis $[P, T_c, \sqrt{e}\sin\omega, \sqrt{e}\cos\omega, K]$ and impose flat priors on all of these orbital elements.  For the planets that transit, we apply Gaussian priors centered on the orbital period and time of conjunction values derived from the transit data with a width equal to the measured uncertainties on these values.  In cases where we include data from multiple telescopes or where the HIRES data include observations taken prior to the 2004 detector upgrade \citep{2005ApJ...632..638V,2016ApJ...821...89B}, we fit a separate RV zero point and jitter value for each dataset.  We do not include datasets that have fewer than 10 data points.  We also bin each set of radial velocity data in two-hour increments, binning datasets from different telescopes separately.  We define our likelihood function in Equation \ref{eqn:likelihood}, where $\sigma_i$ is the instrumental error, $\sigma_{jit}$ is the stellar jitter, $v$ are the data, and $m$ is the model.  

\begin{equation}
\label{eqn:likelihood}
\mathscr{L} = \prod_{i} \frac{1}{\sqrt{2\pi}\sqrt{\sigma_i^2 + \sigma_{jit}^2}}\exp\bigg(-0.5\bigg(\frac{(v_i - m_i)^2}{\sigma_i^2 + \sigma_{jit}^2}\bigg)\bigg)
\end{equation}

We initialize each MCMC chain using the best-fit parameters from our fit.  We note that for several systems we fit a different number of Keplerian orbits than the published number of planets (Table 1).  Of the transiting planet systems, this includes Kepler-20 and Kepler-407.  For Kepler-20 there are six published planets but we fit four Keplerian orbits, as two of the transiting planets did not yield statistically significant RV semi-amplitudes in previous studies \citep{2016AJ....152..160B}.  For Kepler-407, there is one transiting planet and a long-term trend with curvature that was published as a planet detection, but we only fit a full Keplerian orbit for the inner planet as the outer planet's orbital period is poorly constrained by the current RV data \citep{2014ApJS..210...20M}. 

For the RV-detected planetary systems, we search for periodic signals in the radial velocity datasets using the automated planet search pipeline described in \citet{2015ApJ...805..175F}.  We fit Keplerian orbits to all signals with empirical false alarm probabilities (eFAP) less than or equal to 1$\%$ and $K$ greater than or equal to 1.0 m/s in our final RV analysis.  For systems GJ 667, GJ 876, GJ 581, and HD 40307, we find that we are only able to recover a subset of the previously published planets, and fit for 2, 3, 2, and 4 planetary orbits, respectively in these systems. 

For GJ 3341, HD 156668, HD 175607, and GJ 163 we find additional periodic signals with eFAP$\leq$1$\%$ that do not correspond to the periods of the confirmed planets in these systems and may be due to either stellar activity or additional unconfirmed planetary companions. We determine whether to include these additional periodic signals by comparing model fits with and without these additional signals using the Bayesian Information Criterion (BIC).  The BIC is defined as:  $\mathrm{BIC} = -2L + k\ln{n}$, where $L$ is the log likelihood of a model fit, $k$ is the number of free parameters in the model, and $n$ is the number of data points.  In this case, the preferred model is the one with the lowest BIC value.  If the BIC value for the model with additional periodic signals is smaller than the BIC value for the model without the additional periodic signals by at least 10 \citep[a reasonable rule of thumb for statistically significant improvements in fit;][]{doi:10.1080/01621459.1995.10476572}, we consider the model with additional periodic signals to be a better fit, and include these signals in subsequent analyses.   

For GJ 3341, using the automated planet search pipeline we recover the known 14.2 day period planet and also detect a second signal with a period of 202 days and an amplitude of 2.0 m/s (eFAP = 1$\%$).  When we compare BIC values between model fits to the RV data with and without this 202 day signal, we find that the BIC value for the model with the additional periodic signal is only slightly smaller than the BIC value for the model without the additional periodic signal ($\Delta$BIC$=3.9$).  We therefore do not include this additional periodic signal in the RV model fits to the data.

For HD 156668, the known planet with a period of 4.6 days is easily detected by our automated pipeline.  We also detect a second signal at a period of 808 days with an amplitude of 2.9 m/s and a very low false alarm probability.  This appears to be a promising planet candidate but will require additional vetting in order to assess its planetary nature. When comparing model fits, we find that $\Delta$BIC = 99.9 between the model without the additional signal and the model with the additional signal, and thus include this 808 day signal in our RV model fits.  If real, this signal would correspond to a planet with $M\sin i$ = 31 M$_{\Earth}$.  

For HD 175607, which has a known planet with an orbital period of 29 days, we detect a second signal with an eFAP of 0.5$\%$ and a period of 707 days.  However, this period is very close to two years and has poor phase coverage as a result.  We also see a third peak in the periodigram at double this period ($\sim$1400 days), indicating that there is some ambiguity in the true period of this signal.  When comparing model fits, we find that $\Delta$BIC = 27.4 between the model without the additional signal and the model with the additional signal, and thus include this 707 day signal in our RV model fits. We note however that this additional signal will likely require additional RV observations to confirm or disprove its planetary nature. If real, this signal would correspond to a planet with $M\sin i$ = 24 M$_{\Earth}$.

For GJ 163 we detect signals corresponding to the three previously confirmed planets as well as two additional signals at periods of 19 and 108 days. \citet{2013A&A...556A.110B} previously identified these two signals as potential planet candidates.  We find that a model including these additional signals is a significantly better fit to the RV data than a model that does not include these signals ($\Delta$BIC = 12.1), so we include all five signals in our RV analysis.  

When comparing orbital solutions for the RV sample, we found that four of these systems had orbital solutions that were inconsistent at the $>$2$\sigma$ level with published values:  61 Vir, GJ 273, GJ 667, and GJ 876.  For 61 Vir the published orbital solution came from fits to 206 RVs with no linear trend included \citep{2010ApJ...708.1366V}, whereas in our fits we use 786 RVs and include a linear trend (significance 2.2$\sigma$).  For GJ 273, the published orbital solution utilizes a dataset of 279 RVs and does not fit a linear trend.  We include an additional 75 data points from \citet{2017AJ....153..208B} in our fits and detect a linear trend with a significance of 19$\sigma$.  For GJ 667, the published solution uses seven Keplerian orbits (five confirmed planets, two additional signals) and a linear trend to fit a total of 214 RV measurements.  In our blind planet search we only recover three of the seven signals using a marginally bigger RV dataset (238 measurements in total) and including a linear trend.  Finally, GJ 876 is a dynamically rich system with three planets in a Laplace resonance.  Previous studies fit the RV data for this star using N-body code with four planets, while we fit these data using Keplerian orbits for the three inner planets (our blind planet search did not identify the fourth planet).

After fitting our model to each data set, we search for systems with statistically significant linear trends (defined here as fits where the linear slope differs from zero by more than $3\sigma$).  We list the best-fit trend values from the maximum-likelihood fit for each system in Table 1, with corresponding uncertainties determined from the MCMC chains.  We find that 14 of the 65 systems in our sample have statistically significant trends.  We used the BIC to determine whether these statistically significant long term trends were best modeled with a linear trend, a quadratic trend, or an additional Keplerian orbit.  In all cases but one (GJ 676) we found that a linear trend was the preferred model.  For GJ 676, the curvature of the trend was significant enough to justify a fit with a full Keplerian orbital model. 

We also find nine systems in our sample with fully resolved outer gas giant companions that were previously identified in the published literature.  For the purposes of this study, we define an outer ``gas giant" as a companion with a mass greater than 0.5 M$_{\rm Jup}$ outside of 1 AU.  Although some super-Earths in our sample also have outer companions with masses smaller than this cutoff, these planets were likely too small to open a gap in the protoplanetary gas disk \citep[e.g.,][]{1986ApJ...309..846L,2006Icar..181..587C,2012ARA&A..50..211K}.  The 1 AU cutoff includes all gas giants $>$0.5 M$_{\rm Jup}$ orbiting exterior to lower mass planets, and excludes four gas giant planets orbiting at smaller separations (55 Cnc b at 0.11 AU, GJ 876 b and c at 0.21 and 0.13 AU respectively, and WASP-47 b at 0.05 AU).

We list the properties of these previously confirmed outer gas giant planets in Table 3.  

\begin{deluxetable}{lccc}
\tabletypesize{\scriptsize}
\tablecaption{Properties of Outer Gas Giant Companions on Resolved Orbits}
\label{res_companion_table} 
\tablewidth{0pt}
\tablehead{
\colhead{Companion} & \colhead{Mass (M$_{\rm Jup}$)} & \colhead{a (AU)} & \colhead{Ref.}
}
\startdata
Kepler 94 c & 9.84$\pm$0.63 & 1.60$\pm$0.04 & \citet{2014ApJS..210...20M} \\
Kepler 454 c & 4.46$\pm$0.12 & 1.29$\pm$0.02 & \citet{2016ApJ...816...95G}\\
Kepler 68 d & 0.84$\pm$0.05 & 1.47$\pm$0.03 & \citet{2014ApJS..210...20M}\\
Kepler 48 e & 2.07$\pm$0.08 & 1.85$\pm$0.04 & \citet{2014ApJS..210...20M}\\
55 Cnc d & 3.88$\pm$0.07 &  5.50$\pm$0.03 & \citet{2014MNRAS.441..442N} \\
GJ 832 b & 0.68$\pm$0.09 & 3.56$\pm$0.28 & \citet{2014ApJ...791..114W}\\
HD 181433 c & 0.65 & 1.76 & Bouchy et al (2009)\\
HD 181433 d & 0.54 & 3.00 & Bouchy et al (2009)\\
GJ 676 b & 4.96$\pm$0.96 & 1.82$\pm$0.06 & \citet{2017AJ....153..136S}\\
WASP 47 c& 1.29$\pm$0.06 & 1.38$\pm$0.02 & \citet{2017AJ....153...70S}
\enddata
\end{deluxetable}

\subsection{AO Imaging}

For the systems with statistically significant trends, we obtained AO imaging data to determine whether these systems had stellar companions that might have caused the observed trend.  We identified published AO images \citep{2010PASP..122.1195T,2014A&A...567A.112A,2014ApJ...794...51H,2015MNRAS.449.3160R,2016A&A...595A..77S,2017AJ....153...71F,2017AJ....153..242N,2017AJ....154..122C,2016AJ....152....8K} for all but three of these systems. Of the remaining three systems, two (HD 40307 and HD 85512) had unpublished archival data obtained with the NACO instrument \citep{2003SPIE.4841..944L,2003SPIE.4839..140R} on the Very Large Telescope (VLT).  The HD 40307 data were taken in $Ks$-band with a total integration time of 1.1 hr (ID: 088.C-0832(A), PI: Loehne). The HD 85512 data were obtained in $Ks$-band with a total integration time of 9 minutes (ID: 090.C-0125(A), PI: Mugrauer). Both datasets were obtained without a coronagraph, using a 4-point dither pattern. 

We downloaded the data for both stars from the ESO archive and processed them using the pipeline outlined in \citet{2014ApJ...780...17M}. We did not detect any stellar companions in either of these datasets. We show 5$\sigma$ Ks contrast values for both systems in Table 4.

\begin{deluxetable}{lcc}
\tabletypesize{\scriptsize}
\tablecaption{5$\sigma$ Contrast Curves}
\tablewidth{0pt}
\tablehead{
\colhead{System} & \colhead{Separation (arcsec)} & \colhead{5$\sigma$ Contrast (mag)} 
}
\startdata
HD 40307 & 0.79	& 9.47	\\
&1.90	& 10.87	\\
&2.99	& 11.72	\\
&4.07	& 11.45	\\
&5.16	& 11.91	\\
&6.24	& 11.90	\\
&7.33	& 12.06	\\
&8.42	& 12.63	\\
&9.50	& 12.66	\\
HD 85512 & 0.79	& 4.41	\\
&1.28	& 6.95	\\
&1.76	& 7.40	\\
&2.25	& 8.06	\\
&2.74	& 8.68	\\
&3.23	& 9.34	\\
&3.69	& 9.95	\\
&4.18 & 9.44	\\
&4.67	& 10.07 \\
GJ 3634 &0.09  &   0.014       \\
     & 0.22     &  3.92       \\
     & 0.35     &  4.81       \\
     & 0.49     &  5.24       \\
     & 0.63     &  6.18       \\
     & 0.77     &  6.71       \\
     & 0.91     &  6.92       \\
     &  1.05     &  7.15       \\
     &  1.18     &  7.11       \\
     &  1.32     &  7.18       \\
     &  1.46     &  7.20       \\
     &  1.60     &  7.11       \\
     &  1.74     &  7.14       \\
     &  1.88     &  7.05       \\
     &  2.02     &  7.05       \\
     &  2.16     &  7.05       \\
     &  2.30     &  7.10      \\
     &  2.43     &  7.05       \\
     &  2.57     &  7.04       \\
     &  2.71     &  7.11       \\
     &  2.85    &  7.05       \\
     &  2.99     &  7.13       \\
     &  3.13     &  7.06       \\
     &  3.26     &  7.01       
\enddata
\end{deluxetable}

For the remaining system (GJ 3634), we obtained $K_c$-band AO images using NIRC2 at Keck on UT Feb 5 2018 with an effective integration time of 9 seconds and a three-point dither pattern.  We identified a close pair of candidate companions at a separation of $1^{\prime\prime}.8$ and used a multi-peak point spread function (PSF) to simultaneously fit GJ 3634 and the two candidate companions in each frame where the companion is resolved. We constructed the PSF as a sum of Moffat and Gaussian functions and fit over a circular aperture of 10 pixels in radius, corresponding to twice the full width at half maximum of the PSF as described in \citet{2015ApJ...800..138N}. 

We next integrated the best-fit PSFs for GJ 3634 and its candidate companions over the same aperture to determine their flux ratios. We similarly measured the companion separation and position angle by calculating the difference between the centroids of each star. We then applied the NIRC2 astrometric corrections from \citet{2016PASP..128i5004S} to compensate for the NIRC2 array's distortion and rotation. We find that the easternmost candidate companion (labeled as cc1 in Figure 2) has a flux ratio of 116, corresponding to $\Delta K_c = 5.16$. This companion is separated from GJ 3634 by $1^{\prime\prime}.778 \pm 0^{\prime\prime}.002$ at a position angle of $177.37^{\circ} \pm 0.04^{\circ}$ east of north. For the western candidate companion (cc2), we measure a flux ratio of $75 \pm 8$, corresponding to $\Delta K_c = 4.7 \pm 0.1$. This companion is separated from GJ 3634 by $1^{\prime\prime}.860 \pm 0.^{\prime\prime}002$ at a position angle of $203.40^{\circ} \pm 0.05^{\circ}$ east of north. We calculate our uncertainties as the quadrature sum of measurement uncertainties and the uncertainty in the distortion solution. However, in one of the three frames cc1 did not have a regular PSF shape and as a result we were unable to fit for its peak and location.  With just two independent measurements for this companion, we are unable to calculate an empirical measurement uncertainty and therefore only report the astrometric distortion solution uncertainties.  

For each companion in GJ 3634, we determine stellar masses using PHOENIX spectral models and the \citet{1998A&A...337..403B} zero age main sequence
models.  We first select a PHOENIX model for the primary star based on the published stellar properties and then determine the companion's effective temperature by identifying the PHOENIX model that most closely matches the observed flux ratio.  For each PHOENIX model, we determine the corresponding stellar mass and radius using the Baraffe et al. models.  For both companions, we find best-fit masses of 0.08 M$_{\odot}$.  However, with only one epoch of data for this system, it is not possible to determine whether or not these companions are bound to the primary.  We note, however, that this is a high proper motion target (-566.861, -91.371 mas/yr, \citep{2018yCat.1345....0G}) and astrometric measurements with just a one year baseline would easily determine whether these companions are bound.  We thus examined archival VLT/SPHERE images taken May 2017 and found the companion stars at similar positions as in the Keck images from 2018.  The differences in relative separation of the companions between these two epochs is 30 mas for cc1 and 20 mas for cc2, which is significantly less than the reported proper motion of GJ 3634, allowing us to rule out the case that the pair of companions are distant background stars. 

\begin{figure}[h]
\centering
\includegraphics[width=0.45\textwidth]{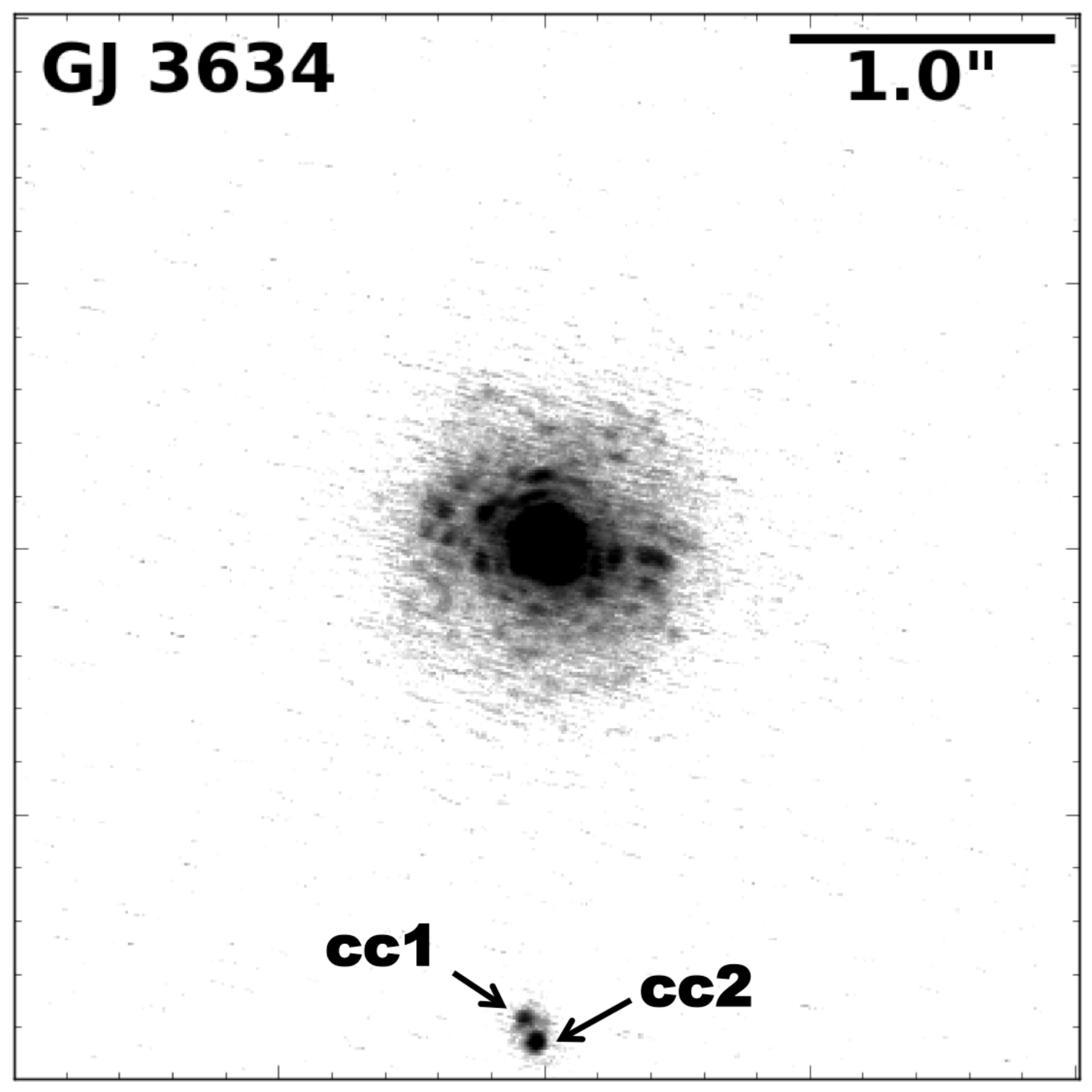}
\caption{Reduced Keck/NIRC2 $K_c$-band image of GJ 3634 showing two candidate companions, labeled cc1 and cc2. Note the image is shown on a log scale, and is aligned with north towards the top and east towards the left.}
\end{figure}

We next consider whether or not the RV trends in these systems might plausibly be explained by the presence of a nearby stellar companion.  Kepler-93, Kepler-97, Kepler-407, and GJ 3634 each have candidate stellar companions, meaning that these systems have one epoch of data showing nearby stars that could be either bound companions or distant background stars. GJ 15A and GJ 676 have confirmed stellar companions that have been shown to have the same proper motion as the primary.  We calculate the minimum companion mass in each system needed to explain the observed RV trend using the equation from \citet{1999PASP..111..169T}:

\begin{align}
M_{\rm comp} = 5.34 \times 10^{-6} M_{\odot} \bigg(\frac{d}{\rm pc}\frac{\rho}{\rm arcsec}\bigg)^2 \nonumber \\
\times \bigg|\frac{\dot v}{\rm m s^{-1}yr^{-1}}\bigg|F(i,e,\omega,\phi).
\end{align}

In this equation, $d$ is the distance to the star, $\rho$ is the projected separation of the companion and the star on the sky, $\dot v$ is the radial velocity trend, and $F(i,e,\omega,\phi)$ is a variable that depends on the orbital parameters of the companion that are currently unconstrained. We use the minimum value of $F$, $\sqrt{27}/2$, which corresponds to the minimum companion mass given the other variables RV slope, distance, and angular separation \citep{2002ApJ...571..519L}.  We then compare this minimum mass to the estimated mass of the candidate companion, which we calculate using the measured brightness ratio under the assumption that the candidate companion is located at the same distance as the primary star.  We discuss our results for each individual system below.    

Kepler-93 is 96.7 pc away and has a candidate companion with a projected separation of $2^{\prime\prime}.29$ \citep{2016AJ....152....8K}.  With an RV trend of 12.0 m s$^{-1}$ yr$^{-1}$, this trend corresponds to a minimum companion mass of 8.2 M$_{\odot}$.  We estimate the mass of the candidate companion using its measured magnitude M$_{\textnormal{K}}$ = 5.35 and assuming an age of 1 Gyr.  We then use the \citet{1998A&A...337..403B} models to calculate a corresponding mass of 0.57 M$_{\odot}$ for this companion.  This mass is significantly smaller than the minimum mass needed to explain the RV trend, and we therefore conclude that this candidate companion cannot explain the observed RV trend and keep this system in our sample.

Kepler-97 is 414 pc away and has a candidate companion with a projected separation of $0^{\prime\prime}.385$ \citep{2017AJ....153...71F}.  With an RV trend of 4.5 m s$^{-1}$ yr$^{-1}$, this trend corresponds to a minimum companion mass of 1.58 M$_{\odot}$.  The candidate companion in this system has a magnitude M$_{\textnormal{K}}$ = 6.28, corresponding to an estimated companion mass of 0.4 M$_{\odot}$ using its estimated age of 8.4 Gyr.  As this is smaller than the minimum mass needed to explain the RV trend, we leave this system in our sample.

Kepler-407 is 326 pc away and has a candidate companion with a projected separation of $2^{\prime\prime}.13$ \citep{2016AJ....152....8K}.  With an RV trend of -155.8 m s$^{-1}$ yr$^{-1}$, this trend corresponds to a minimum companion mass of 1045 M$_{\odot}$.  Given the companion's measured magnitude of M$_{\textnormal{K}}$ = 7.0 and using its estimated age of 7.5 Gyr, the estimated companion mass is 0.3 M$_{\odot}$.  This is several orders of magnitude smaller than would be required in order to explain the observed RV trend, and we therefore leave this system in the sample.
 
GJ 3634 is 19.8 pc away and has two candidate companions in what appears to be a hierarchical triple system, as discussed above.  These two companions are $1^{\prime\prime}.83$ away from GJ 3634 and have a mutual separation of 0.087$^{\prime\prime}$.  The measured RV trend in this system is 9.6 m s$^{-1}$ yr$^{-1}$, corresponding to a minimum companion mass of 0.018 M$_{\odot}$.  As discussed earlier, both companions have estimated masses of 0.08 M$_{\odot}$, indicating that their combined mass could be responsible for the observed RV trend.  We thus remove this system from our sample in subsequent analyses. We note that the RV trend in this system was previously published in \citet{2011A&A...528A.111B}. Given their trend, they estimate a minimum mass of 32 M$_{\Earth}$ and a minimum period of 200 days.  Our AO image is the first to indicate that this trend might be due to the presence of stellar/brown dwarf companions rather than a distant orbiting planet.

GJ 15A is 3.6 pc away and has a confirmed stellar companion with a projected separation of $20^{\prime\prime}.28$ \citep{2014ApJ...794...51H}.  With an RV trend of -0.44 m s$^{-1}$ yr$^{-1}$, this corresponds to a minimum companion mass of 0.074 M$_{\odot}$.  The stellar companion in this system has an absolute magnitude of M$_{\textnormal{K}}$ = 8.17, corresponding to an estimated companion mass of 0.175 M$_{\odot}$ for an age of 1 Gyr.  As this estimated companion mass is larger than the minimum companion mass needed to account for the trend, we exclude this system from our subsequent analysis.

Finally, GJ 676 is 15.9 pc away \citep{2016A&A...595A...1G,2016A&A...595A...2G} and has a confirmed stellar companion at a separation of $47^{\prime\prime}$.  With an RV trend of 21.6 m s$^{-1}$ yr$^{-1}$, this trend corresponds to a minimum companion mass of 167 M$_{\odot}$. Given an absolute magnitude of M$_{\textnormal{K}}$ = 6.9, the estimated companion mass is 0.3 M$_{\odot}$ assuming an age of of 1 Gyr.  Since this estimated companion mass is well below the minimum mass to account for the observed RV trend, we conclude this companion could not be producing the observed trend and leave this system in our sample.

\subsection{Trends Due to Stellar Activity}
We next consider whether any of the observed trends might be due to stellar activity. We examined each system in order to determine if the measured RV trend exhibits a correlation with the star's emission in Ca II H $\&$ K lines as quantified by either the S$_{\textnormal{HK}}$ index or logR$'$ \citep{2004ApJS..152..261W,2010ApJ...725..875I}. 
We calculated the Spearman-Rank correlation coefficients between the RV data and this activity indicator after subtracting the orbital solutions for the confirmed inner planets.  We considered a correlation coefficient with an absolute value greater than 0.3 to indicate a significant correlation.  We find that systems HD 219134, HD 40307, and HD 85512 have significant correlations between stellar activity and the observed RV trend, and remove these system from our subsequent analysis.  We also remove HD 1461 from our analysis, as we determined in \citet{2016ApJ...821...89B} that this system has a fully resolved long period signal that is significantly correlated with stellar activity.  

There were two systems with RV trends for which we were not able to obtain stellar activity data, including Corot-24 and GJ 3634. We conclude that stellar activity is unlikely to be the cause of the trend in Corot-24, as the amplitude of the observed trend is higher than would be expected for stellar activity signals.  Although we cannot determine whether or not the observed RV trend in the GJ 3634 might be due to stellar activity, we have already removed this system from further analysis due to the presence of candidate stellar companions that could have caused the observed trend.   

After removing systems with either stellar or potentially activity-related sources of RV trends, including HD 219134, HD 85512, HD 40307, HD 1461, GJ 15A, and GJ 3634, we are left with nine systems with statistically significant trends that can plausibly be attributed to the presence of a substellar companion.  We plot the RV data for each of these systems after subtracting the orbital solutions for the confirmed inner planets in Figure 3.  Trends for GJ 667C \citep{2012ApJ...751L..16A}, GJ 676 \citep{2012A&A...548A..58A,2016A&A...595A..77S}, Kepler-93 \citep{2014ApJS..210...20M,2015ApJ...800..135D}, Kepler-97 \citep{2014ApJS..210...20M}, Kepler-407 \citep{2014ApJS..210...20M}, and Kepler-454 \citep{2015ApJ...800..135D} were previously reported in the published literature.

\begin{figure*}
\begin{tabular}{ c  c  c }
\includegraphics[width=0.33\textwidth]{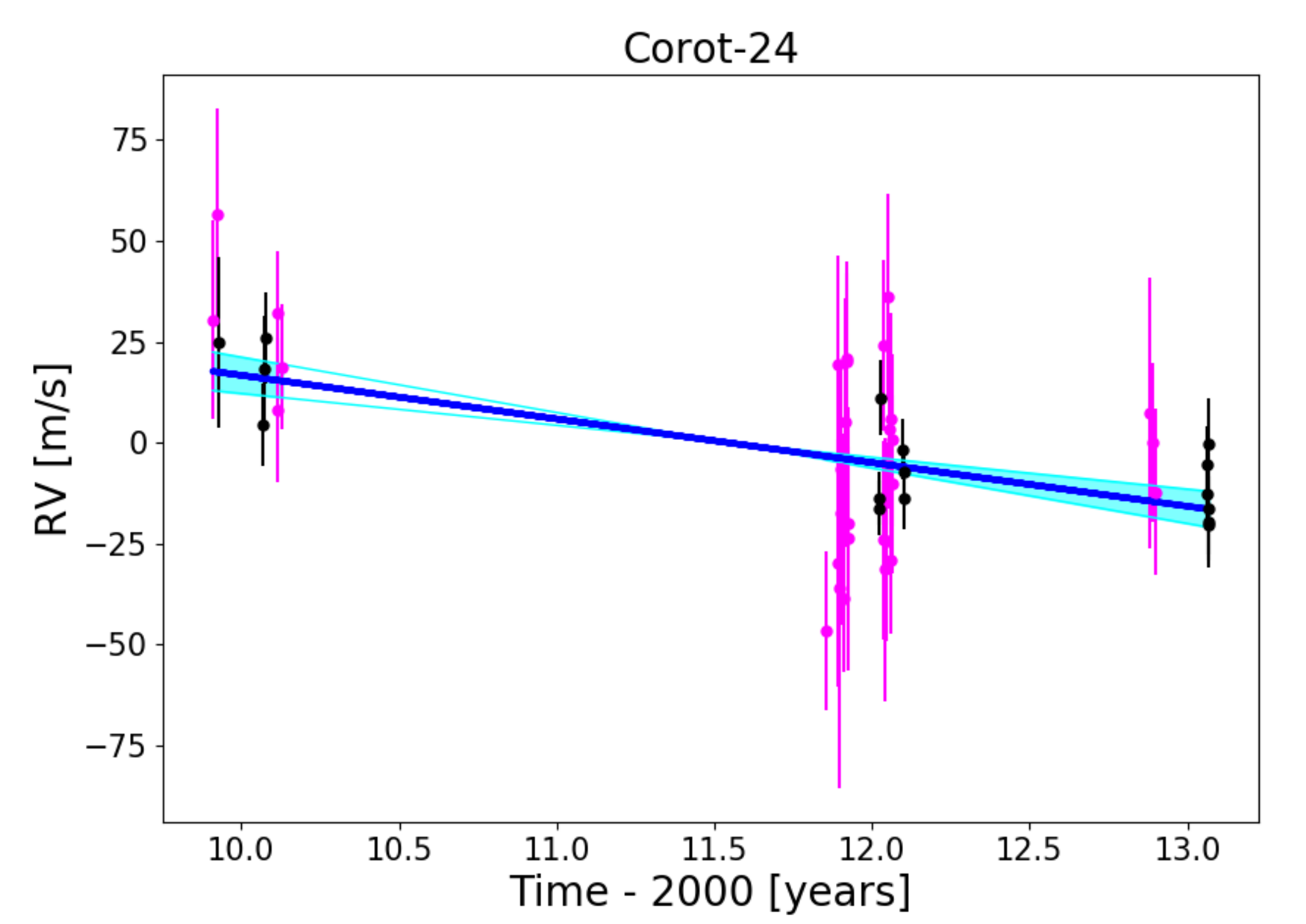} &
\includegraphics[width=0.33\textwidth]{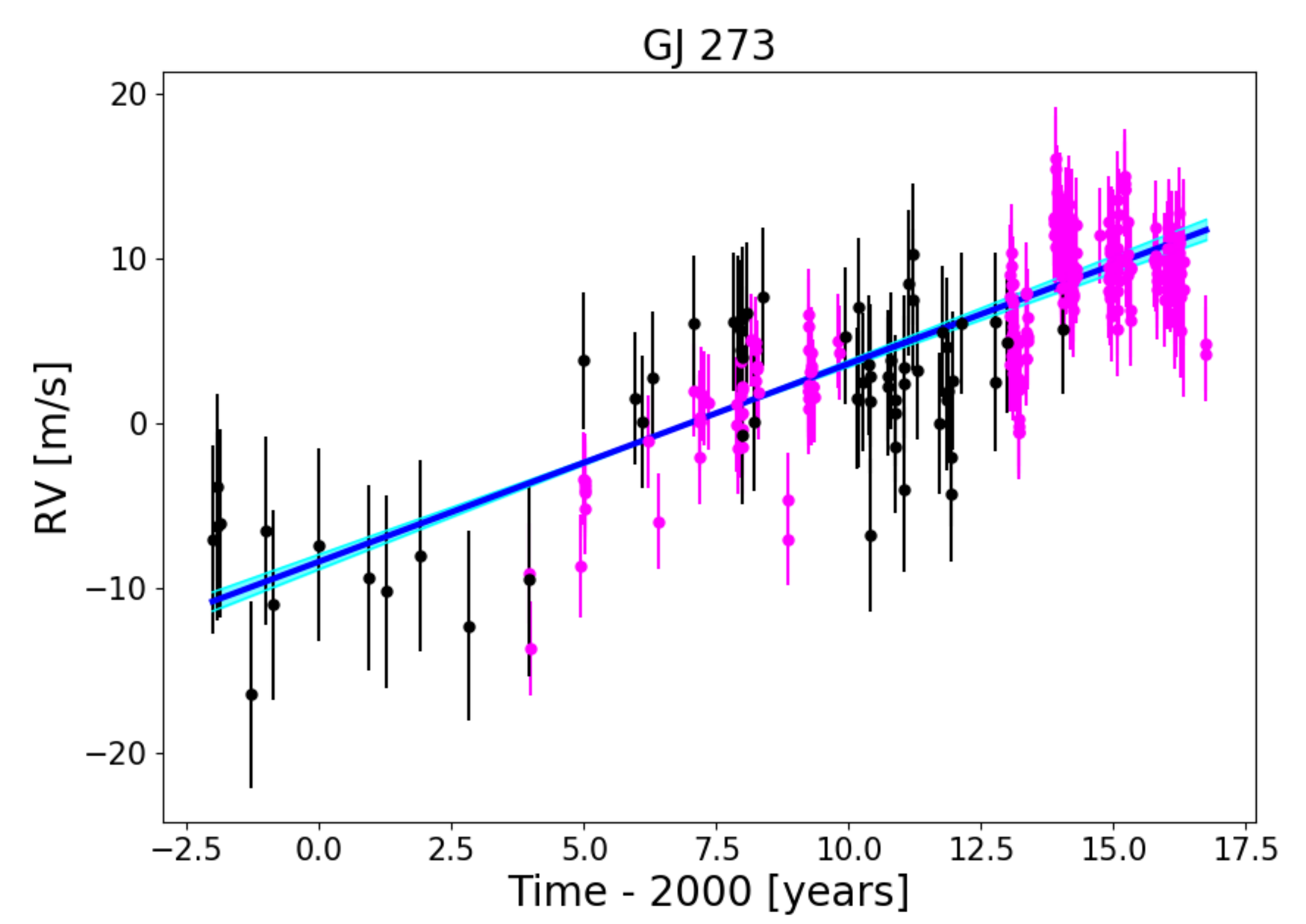} &
\includegraphics[width=0.33\textwidth]{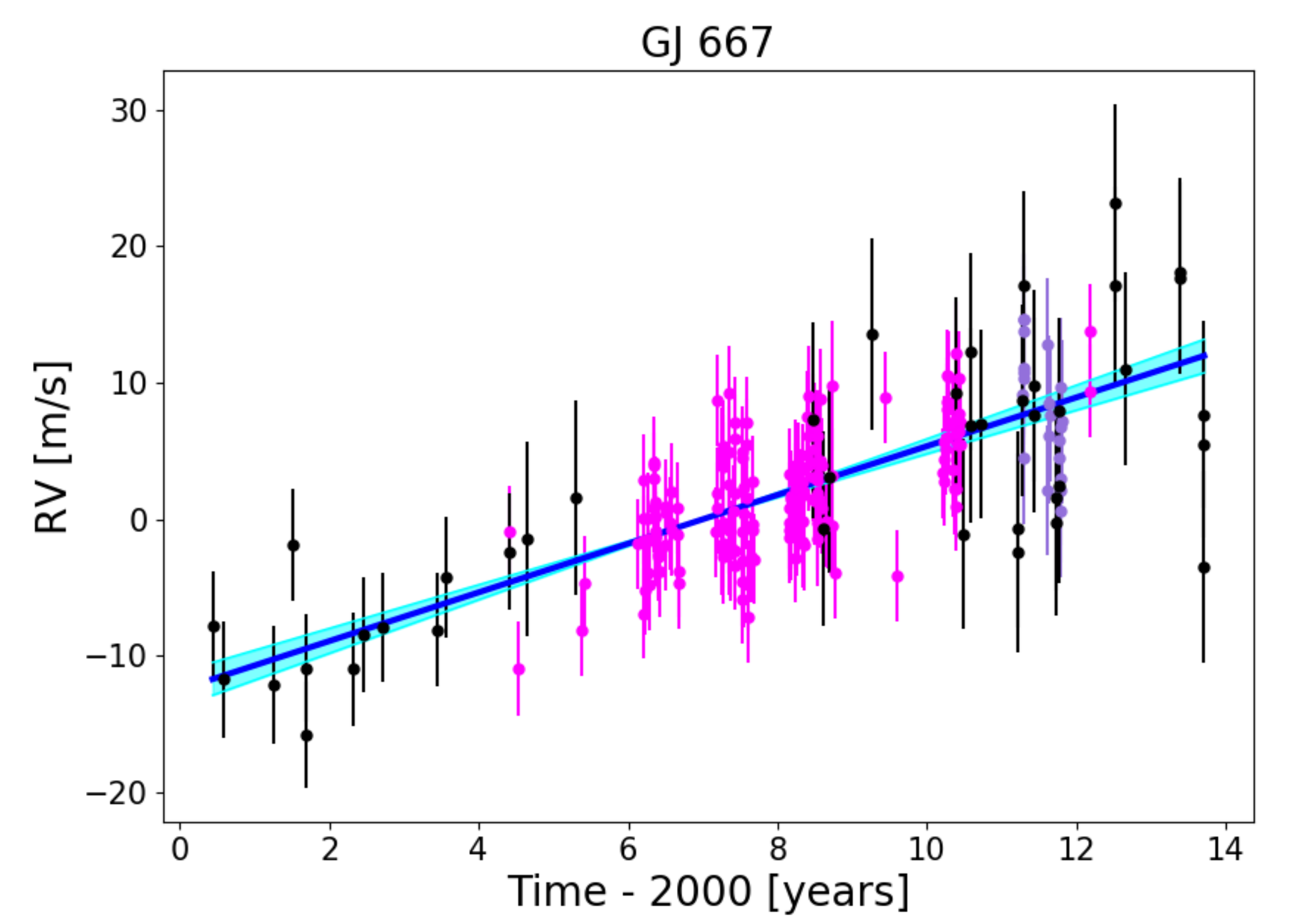} \\
\includegraphics[width=0.33\textwidth]{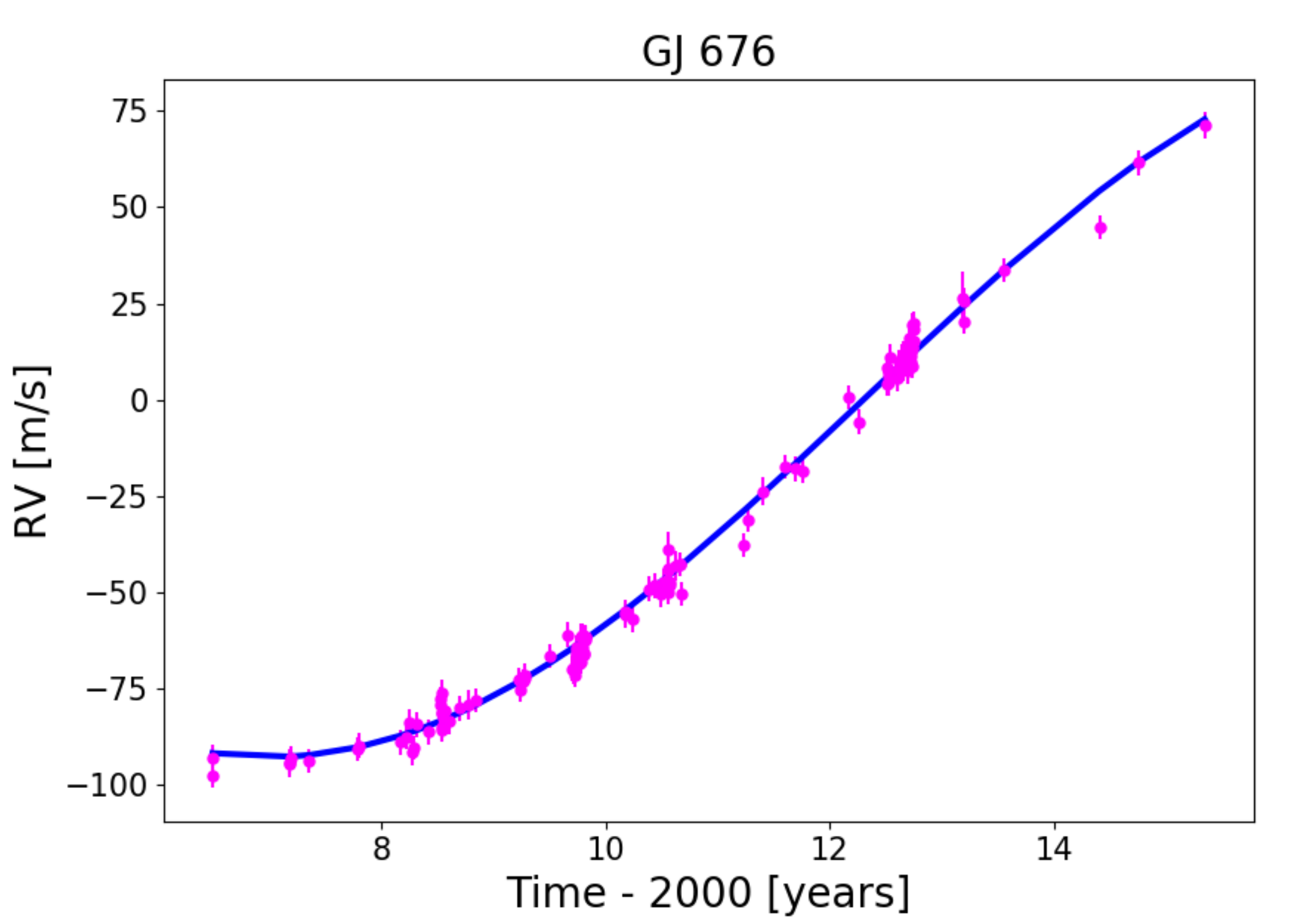} &
\includegraphics[width=0.33\textwidth]{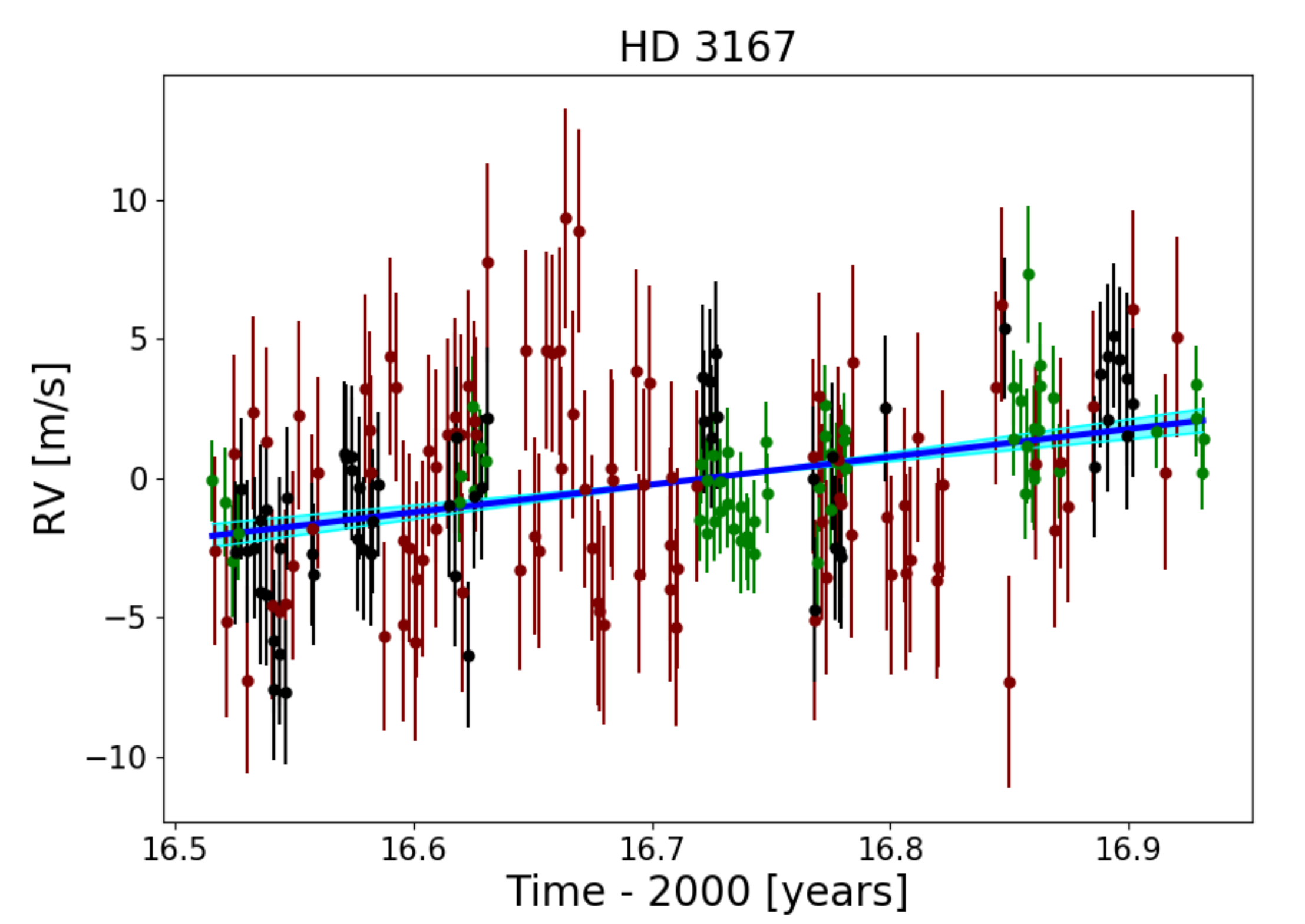} &
\includegraphics[width=0.33\textwidth]{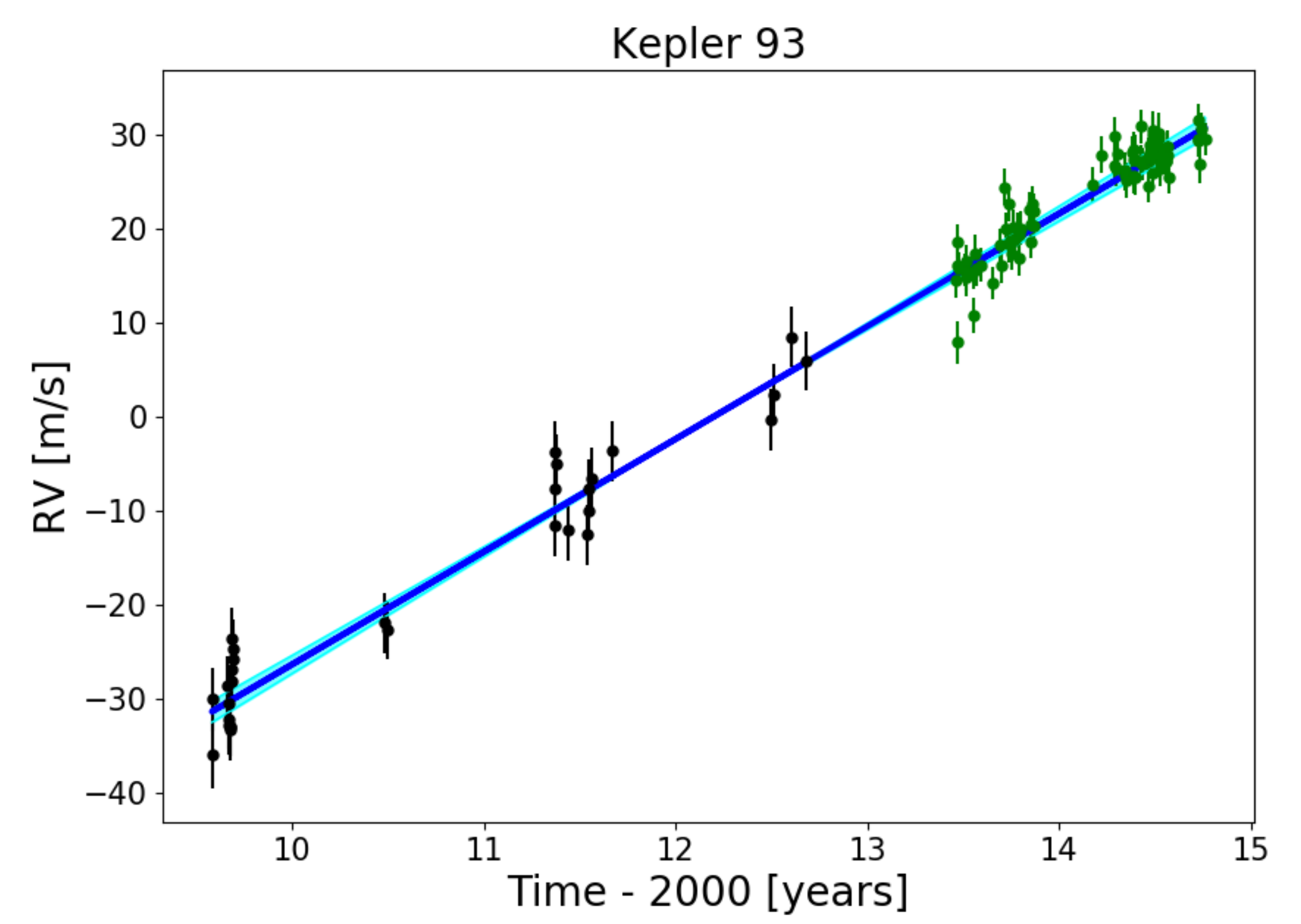} \\
\includegraphics[width=0.33\textwidth]{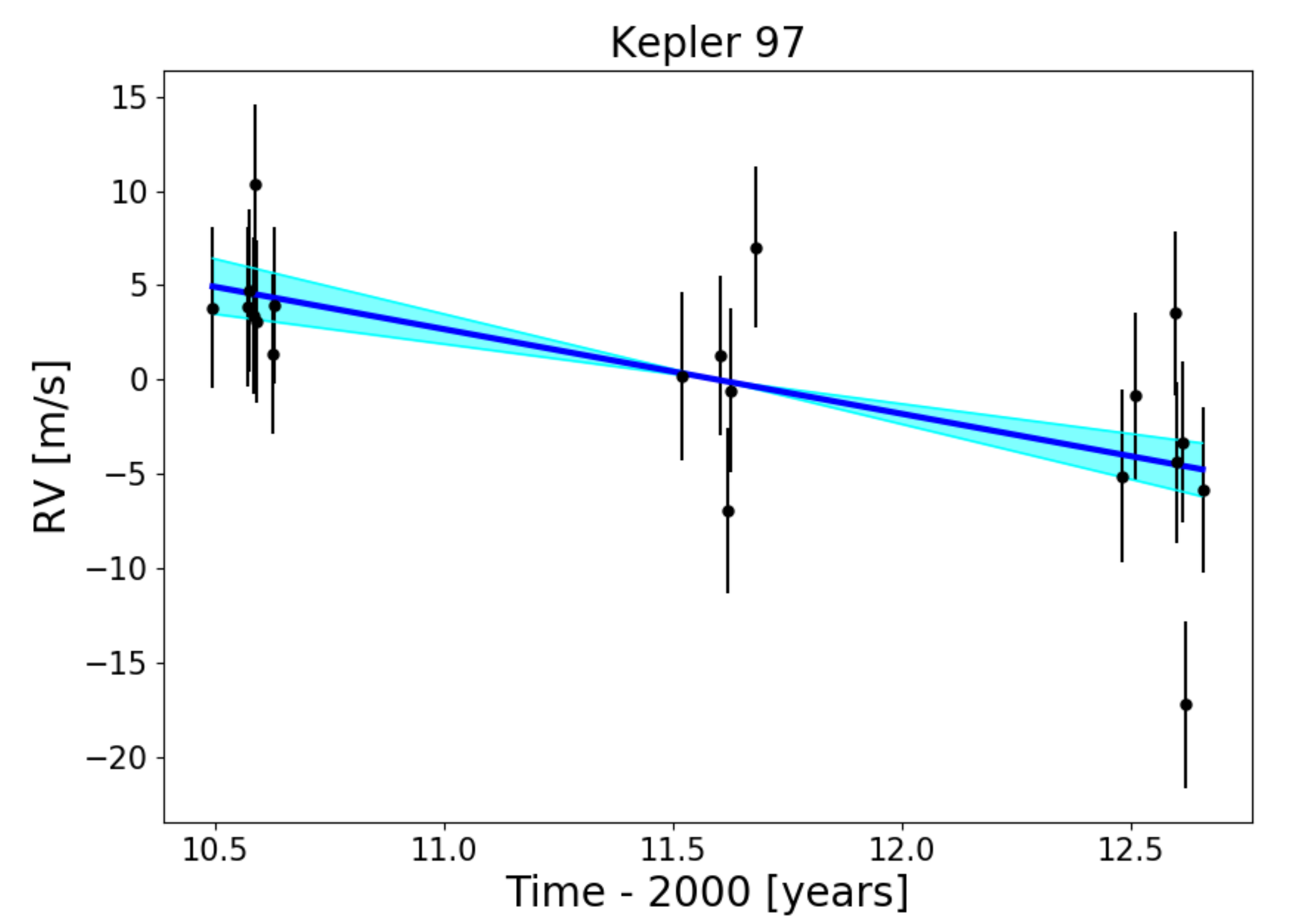} &
\includegraphics[width=0.33\textwidth]{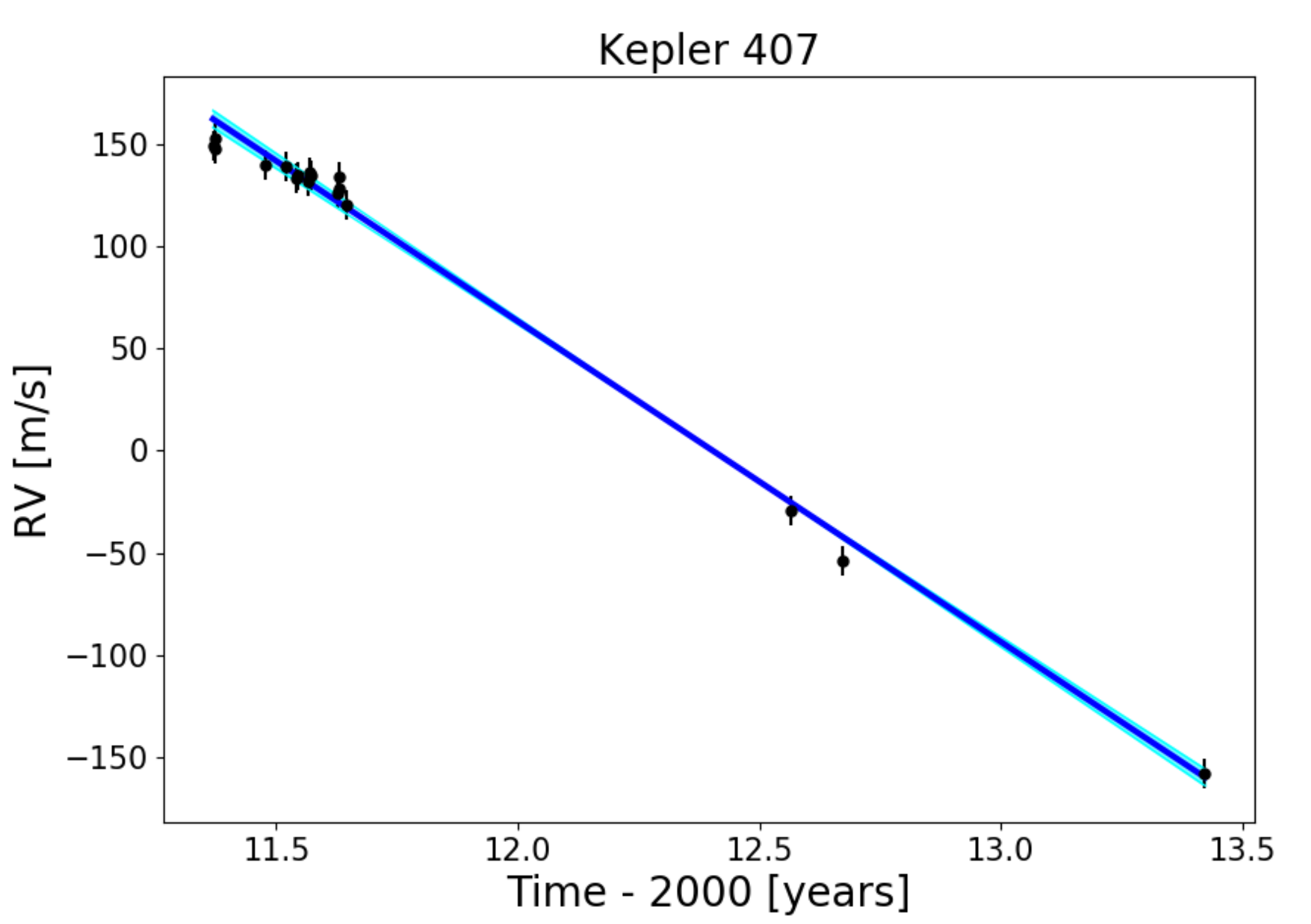} &
\includegraphics[width=0.33\textwidth]{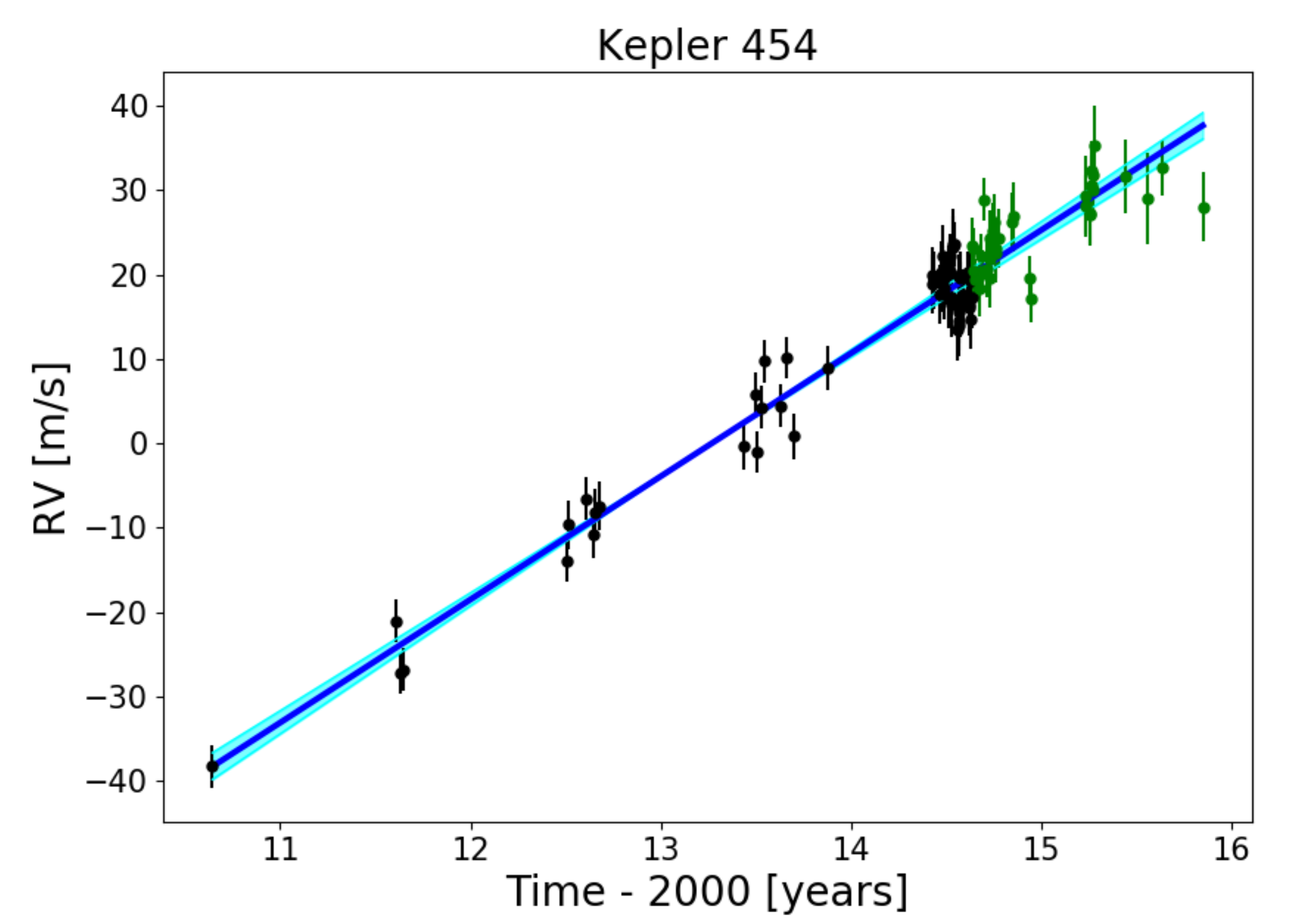}  
\end{tabular}
\caption{Best fit accelerations to the radial velocity data with a $3\sigma$ trend.  The best fit trend is shown as a solid blue line, the $1\sigma$ errors on the slope are presented light blue shaded regions.  The different colored data points represent RVs taken using different telescopes:  black = HIRES, green = HARPS-N, pink = HARPS,  light purple = PFS, maroon = APF. Note that GJ 676 has a curved trend, which allows us to place much tighter constraints on the mass and separation of the companion producing that trend.}  
\end{figure*}

\subsection{Constraints on Companion Masses and Orbital Semi-Major Axes}

We use the RV data to place constraints on the masses and semi-major axes of the long period companions in each system.  The duration and shape of the RV trend places a lower limit on the companion's mass and separation, while the lack of a detection in our AO imaging data places a corresponding upper limit on these quantities.  As described in \citet{2016ApJ...821...89B}, we calculate two-dimensional probability distributions for each companion using an equally spaced 50$\times$50 grid in logarithmic mass (true mass, not $m\sin i$), and logarithmic semi-major axis spanning a range of 0.3 - 500 M$_{\textnormal{Jup}}$ and 0.5 - 500 AU.  In each grid cell we inject 500 simulated companions and determine whether or not they are consistent with the RV observations as follows.  We first draw a set of orbital parameters for the confirmed inner planets from the previous MCMC fits, and then subtract away this orbital solution to preserve any long-term trend signal.  We then draw a mass and semi-major axis value from within the grid box from a uniform distribution in log(M) and log(a), and draw an inclination from a uniform distribution in $\cos i$.  We draw our eccentricity values from a beta distribution with $a = 1.12$ and $b = 3.09$, which are derived from a fit to the population of long-period gas giant planets from RV surveys \citep{2013MNRAS.434L..51K}.  Given a fixed semi-major axis, mass, and eccentricity for each simulated companion, we then fit for the remaining orbital parameters including time of periastron, argument of periastron, and a velocity zero point and calculate the corresponding log likelihood value of the best-fit solution.  

After repeating this process five hundred times in each grid cell, we convert the resulting $50\times50\times500$ cube of log likelihood values to probabilities and marginalize over our 500 samples in each grid cell to yield a two-dimensional probability distribution in mass and semi-major axis for each system.  We calculate two-dimensional probability distributions for all systems in our sample, regardless of whether or not they have statistically significant trends.  The only difference between those systems with and without trends is that we use our AO imaging data to place an upper limit on the companion mass and semi-major axis in the trend systems as discussed in \citet{2016ApJ...821...89B}.  We note that for GJ 273, as a result of its close distance (3.8 pc) and the limited angular extent of the available contrast curve, the contrast curve for this system does not provide significant constraints on the mass and separation of the companion where the probability density for the companion is large.  Similarly, for Corot-24 due to the significant distance of the system (600 pc), constraints provided by the available contrast curve at smaller separations where the probability density for the companion is high are not significant. 

Figure 4 shows the posterior distributions for the nine systems with $3\sigma$ trends, while Table 5 indicates the corresponding $1\sigma$ limits in mass and semi-major axis for each companion.

\begin{figure*}
\begin{tabular}{ c  c c }
\includegraphics[width=0.33\textwidth]{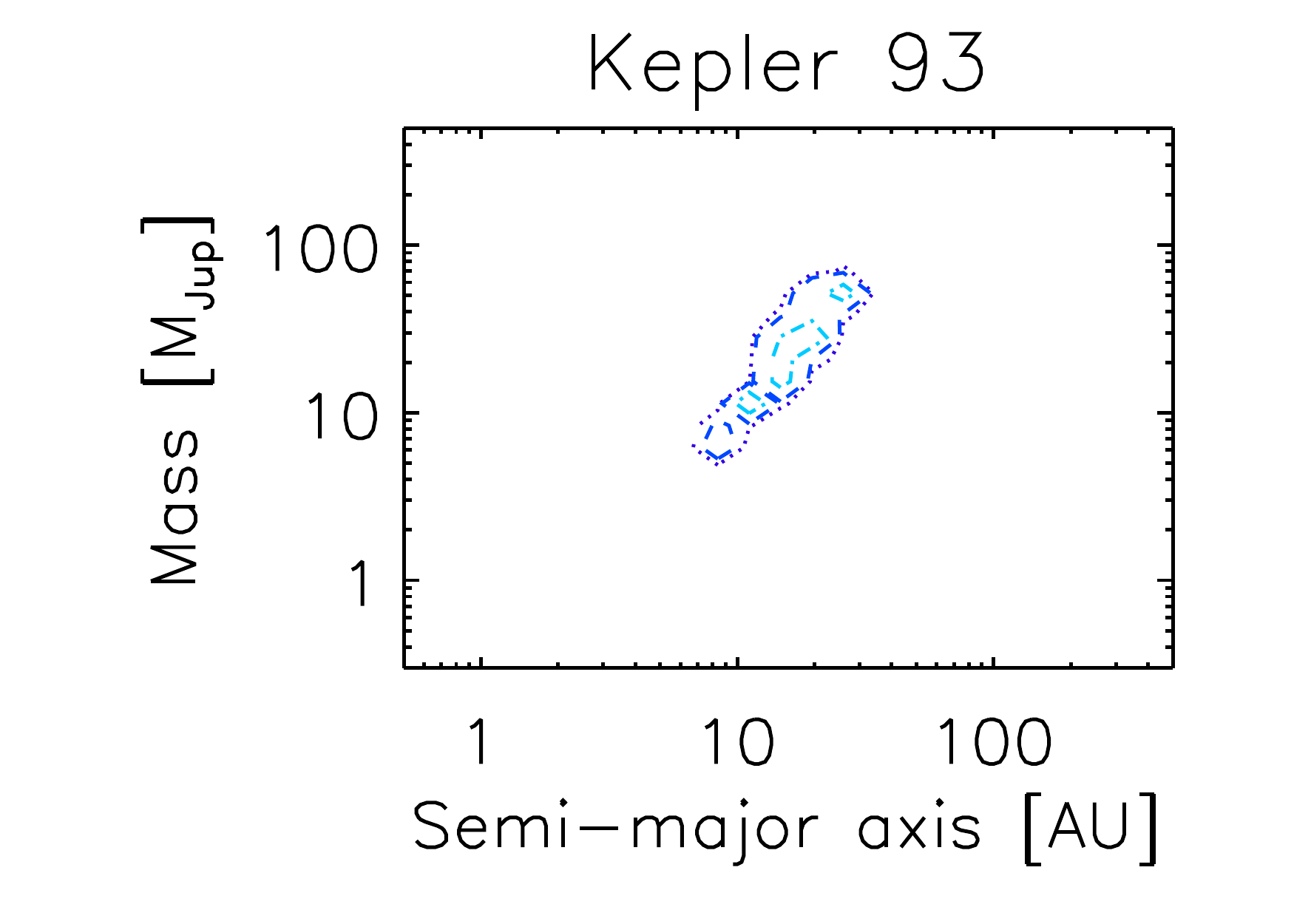} &
\includegraphics[width=0.33\textwidth]{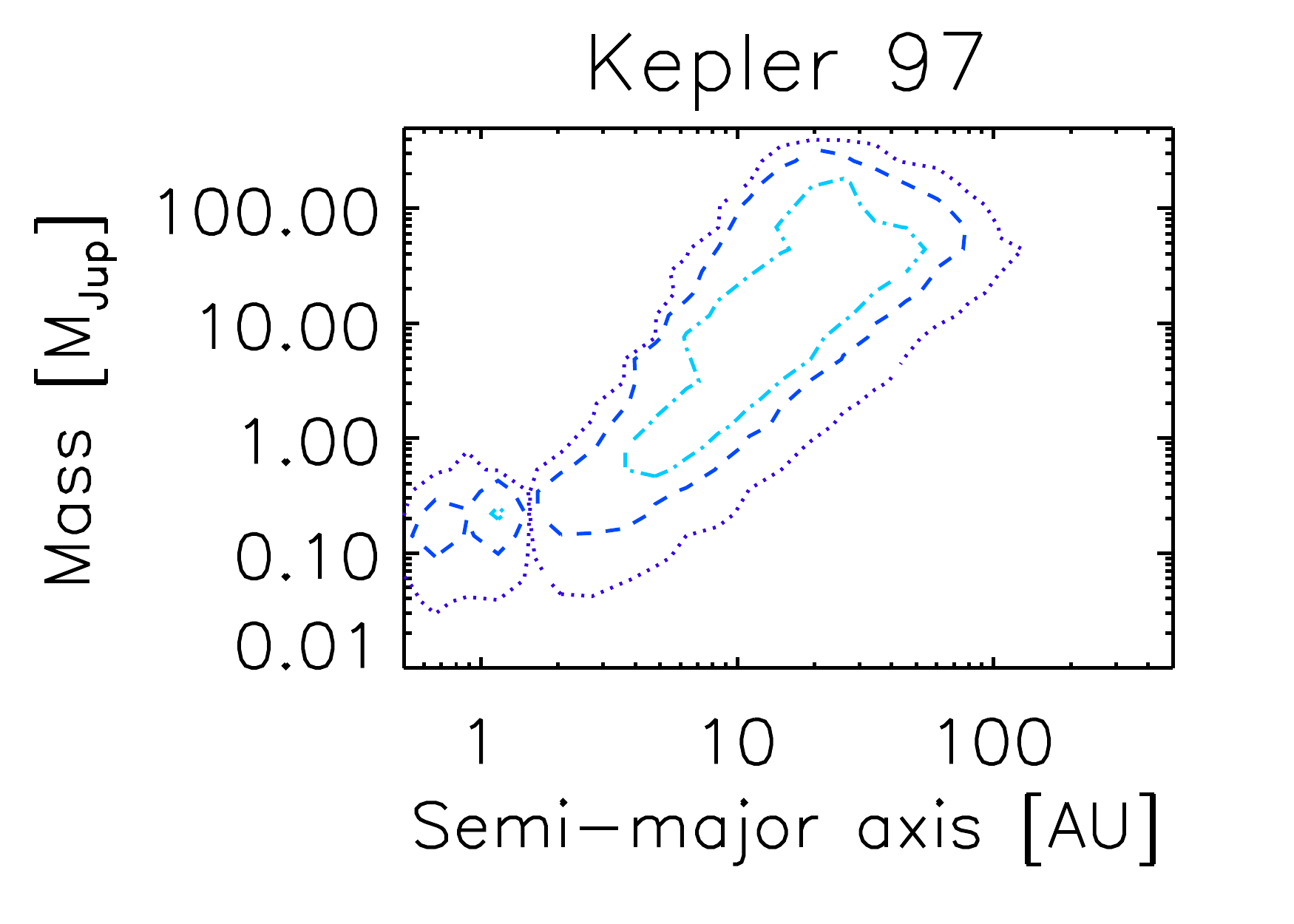} &
\includegraphics[width=0.33\textwidth]{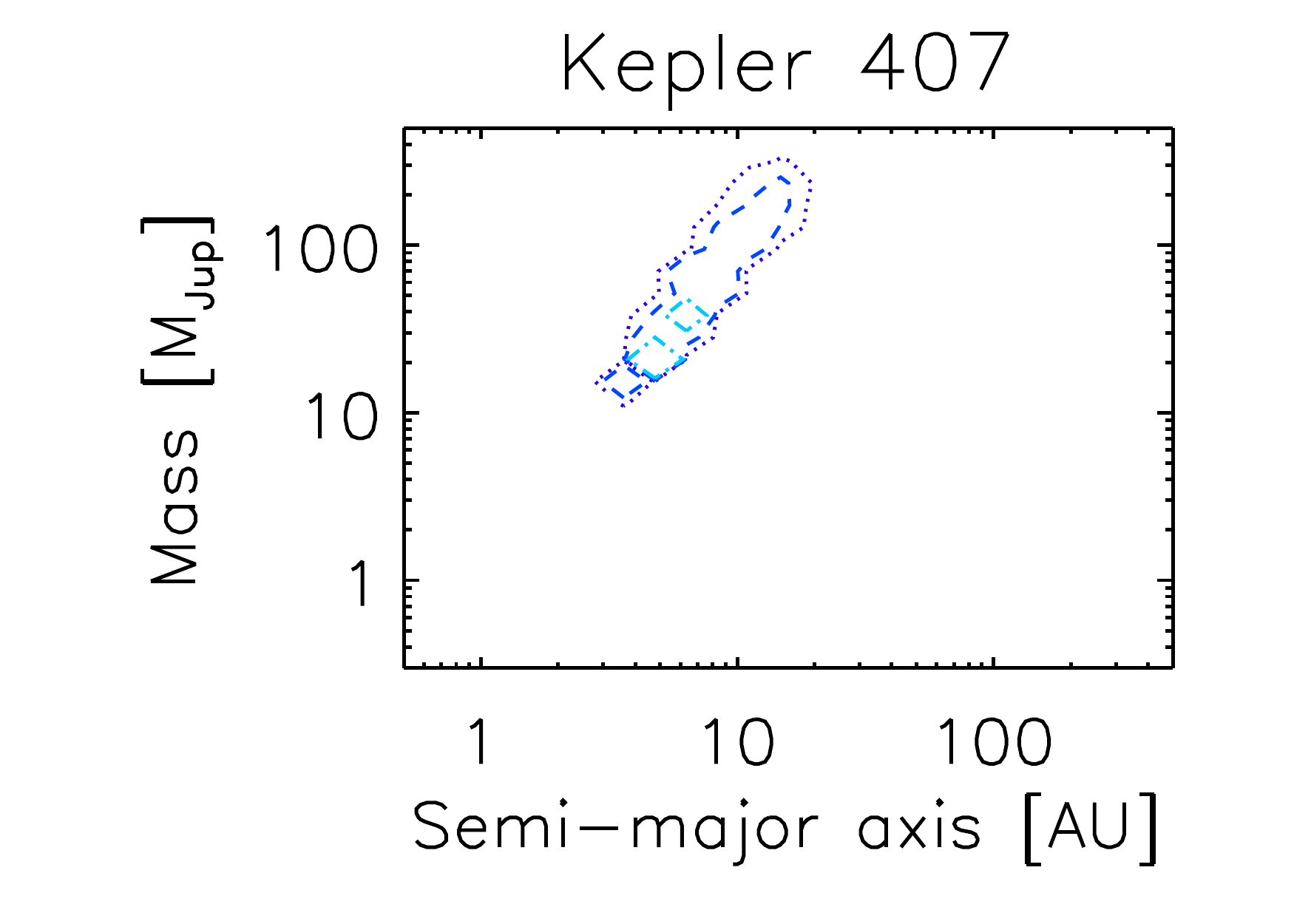} \\
\includegraphics[width=0.33\textwidth]{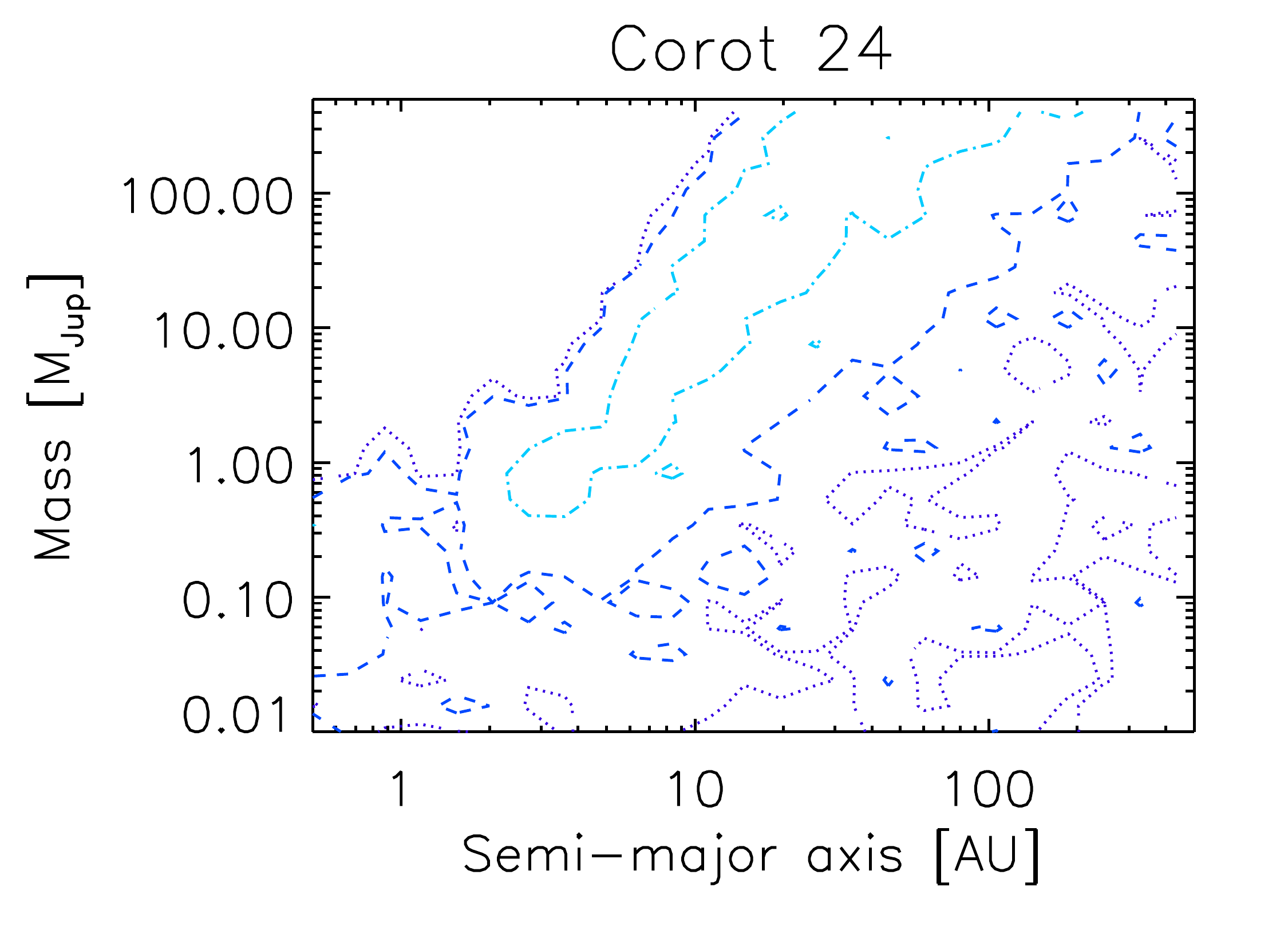} &
\includegraphics[width=0.33\textwidth]{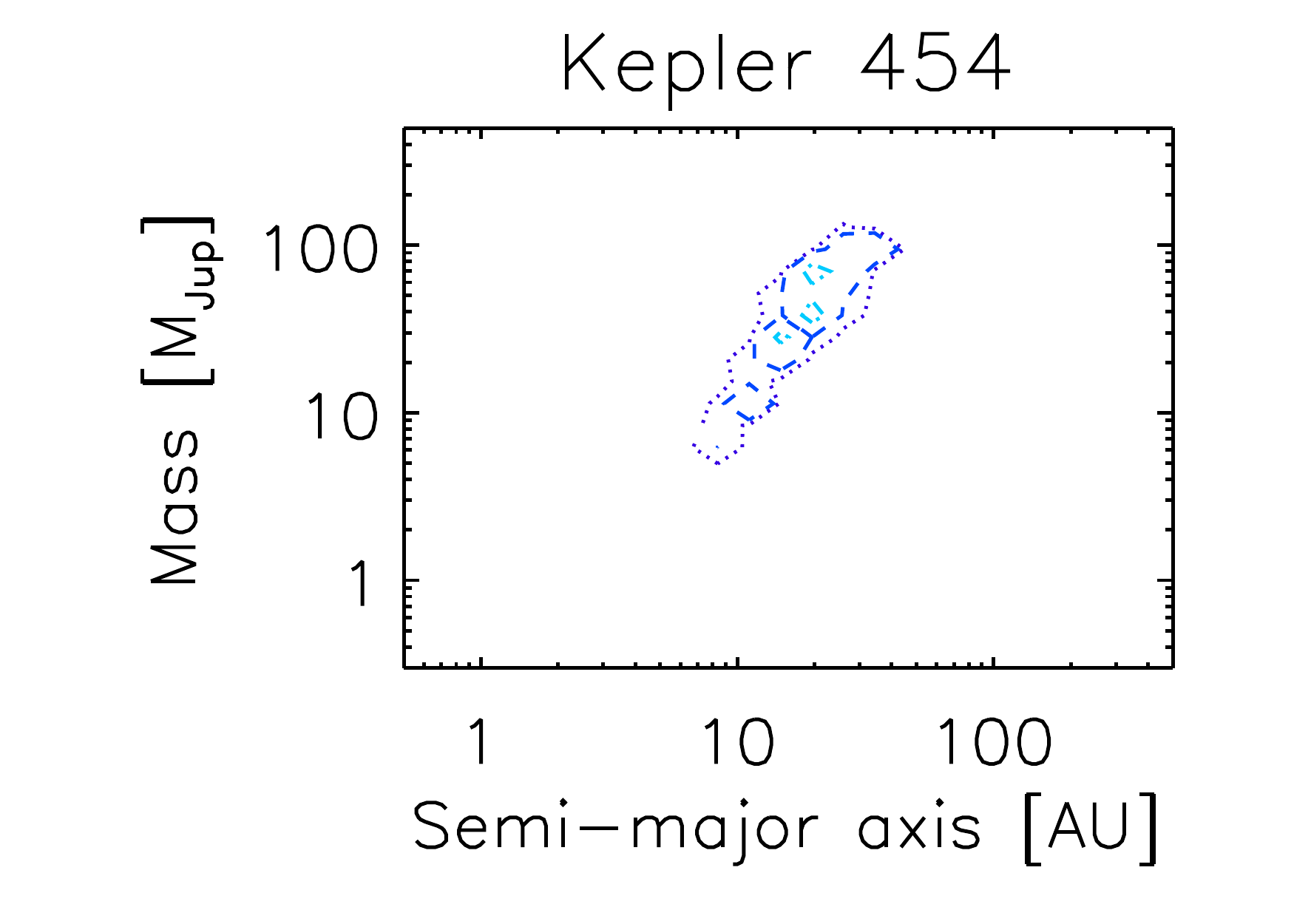} &
\includegraphics[width=0.33\textwidth]{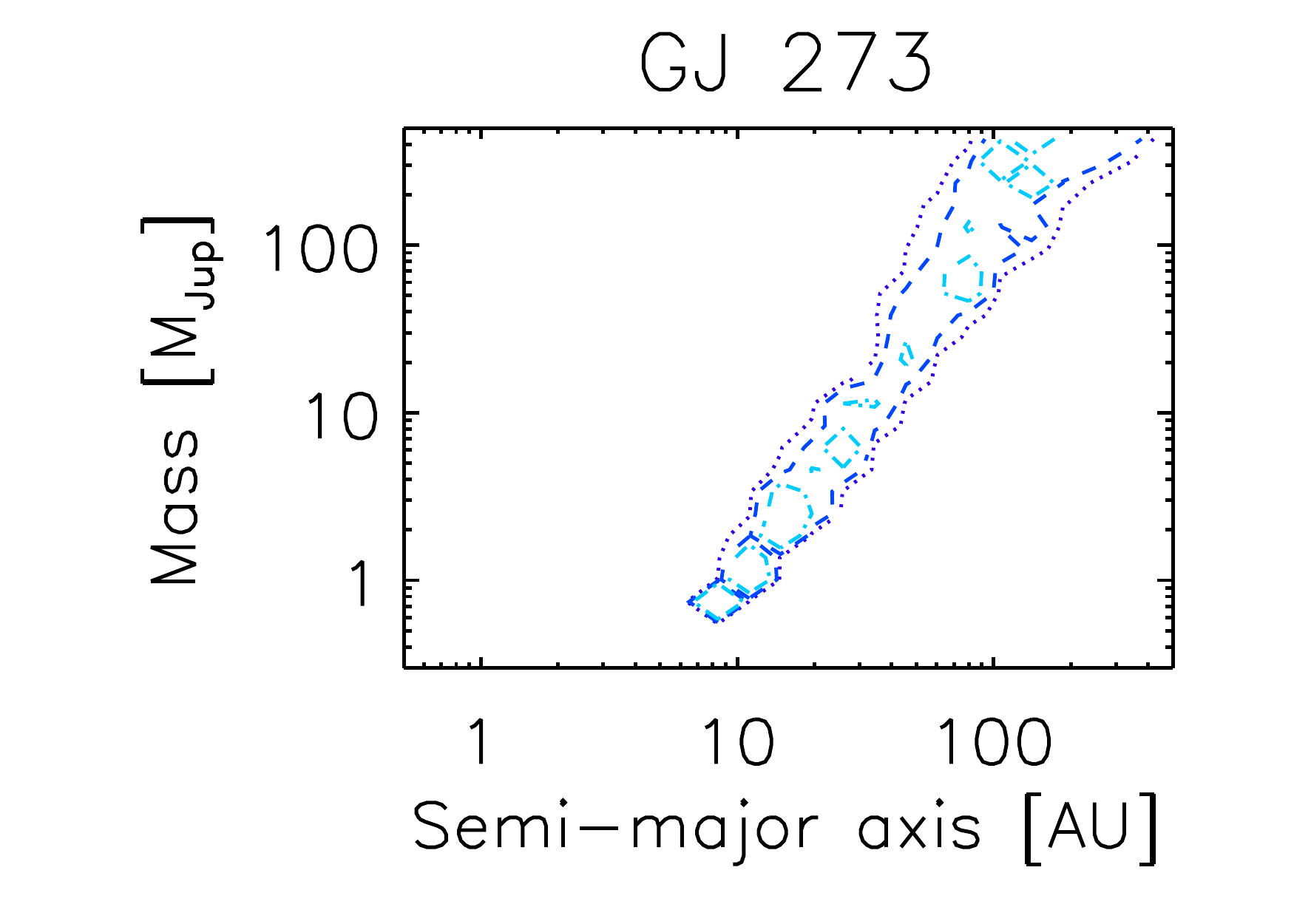} \\
\includegraphics[width=0.33\textwidth]{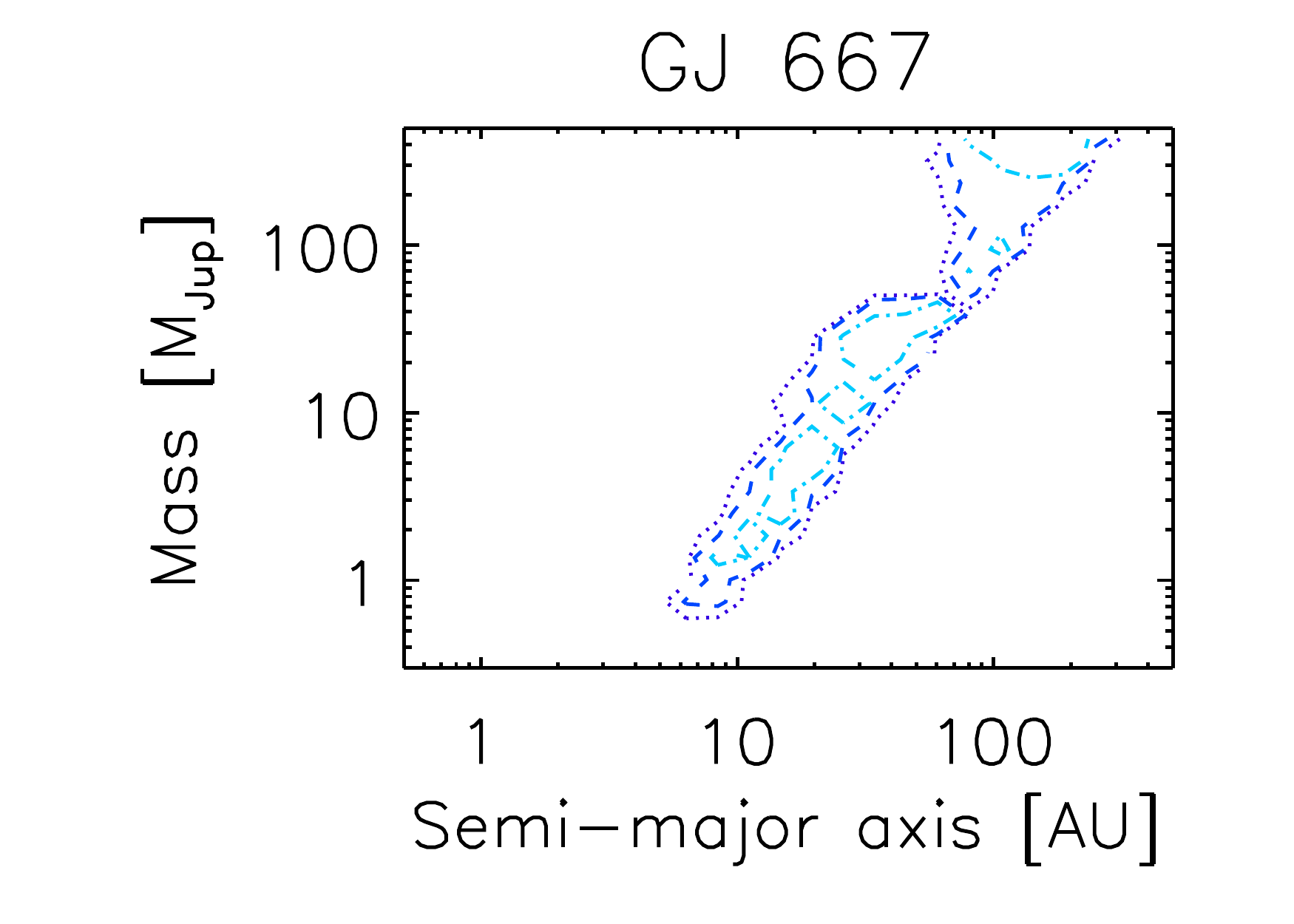} &
\includegraphics[width=0.33\textwidth]{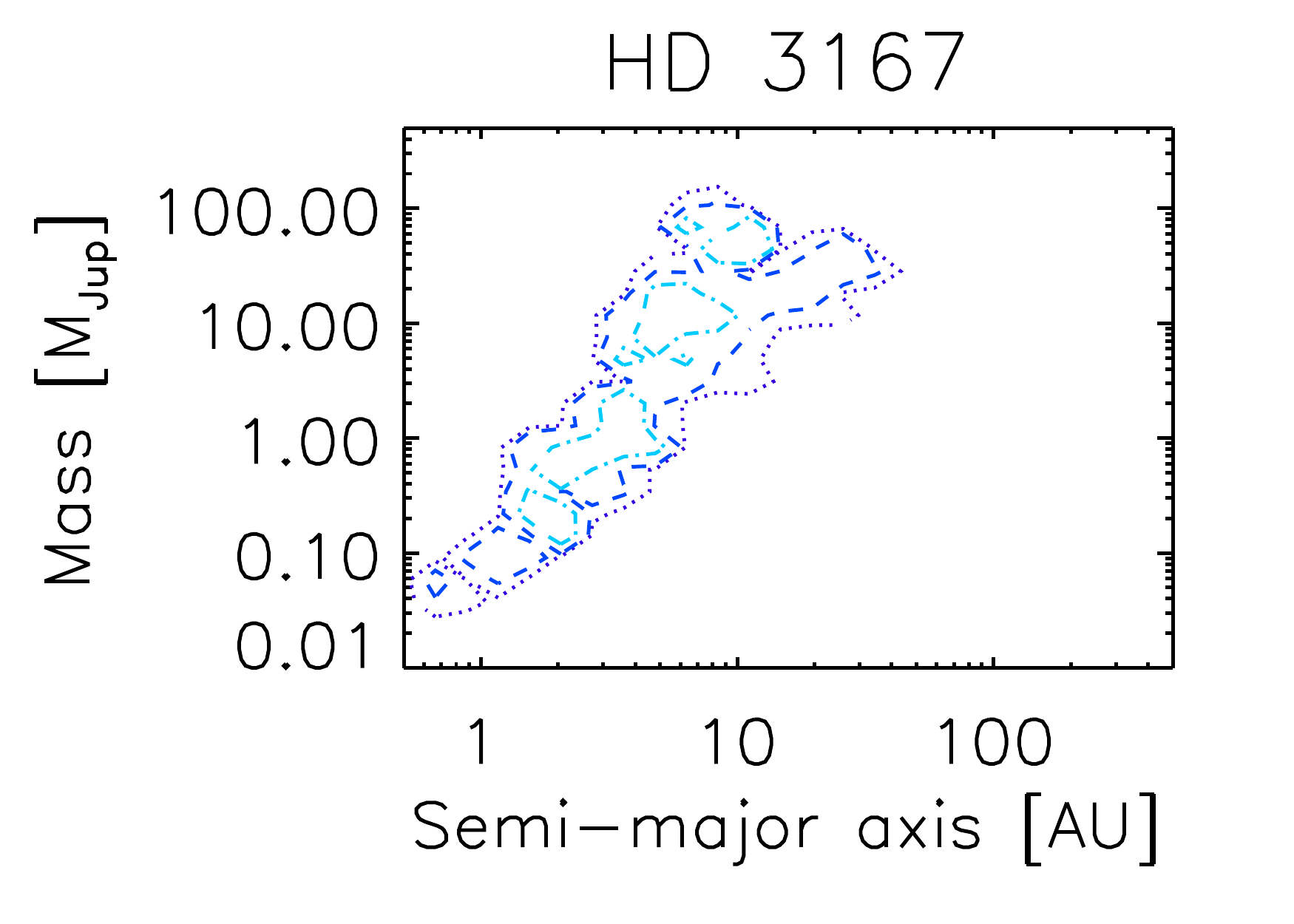} 
\end{tabular}
\caption{Probability distributions for the nine systems with statistically significant trends that are plausibly due to an orbiting substellar companion (i.e., they cannot be explained by either stellar activity or the presence of any known stellar companion).  The three contours define the $1\sigma$, $2\sigma$, and $3\sigma$ levels moving outward.  We do not show the probability distribution for GJ 676 here, as the probability density is concentrated in just a few grid points and the contours are therefore unresolved.}
\end{figure*}

\begin{deluxetable}{lcc}\label{unres_companion_table}
\tabletypesize{\scriptsize}
\tablecaption{Constraints on Companion Properties}
\tablewidth{0pt}
\tablehead{
\colhead{Companion} & \colhead{Mass (M$_{\rm Jup}$)} & \colhead{Semi-major Axis (AU)} 
}
\startdata
Kepler 93 c &11.3 - 51.6 &9.6 - 25.9 \\
Kepler 97 c &0.18 - 166 & 1.2 - 60.3\\
Kepler 407 c &11.4 - 51.6 &3.1 - 7.3 \\
Corot 24 d & 0.27 - 401& 0.5 - 186\\
Kepler 454 d & 7.2 - 81.3 & 9.6 - 29.8\\
GJ 273 d& 0.55 - 430& 7.3 - 214\\
GJ 667 h& 1.2 - 430& 8.4 - 214\\
HD 3167 e&0.05 - 85 & 0.8 - 22 
\enddata
\end{deluxetable}

\subsection{Completeness Maps}

We evaluate our sensitivity to distant companions in each system by calculating the completeness as a function of mass and orbital semi-major axis after taking into account the time baseline, number of data points, and measurement errors for each dataset. As before, we start with a 50$\times$50 grid in mass and semi-major axis evenly spaced in log space from 0.3 - 500 M$_{\textnormal{Jup}}$ and 0.5 - 500 AU.  For each grid box we inject 500 simulated companions where we draw a mass and semi-major axis from a uniformly spaced distribution across each grid box, an eccentricity value from the $\beta$ distribution, inclination from a uniform distribution in $\cos i$, and the remaining orbital elements from uniform distributions. We then calculate the RV signal from this simulated companion at each observation epoch.  We add noise into these simulated RVs by drawing from a Gaussian distribution with a width defined by $\sqrt{\sigma_i^2 + \sigma_{\textnormal{jit}}^2}$, where $\sigma_i$ is the instrumental uncertainty (randomly shuffled from the original dataset) and $\sigma_{\textnormal{jit}}$ is the stellar jitter estimated from the earlier MCMC fits.  To assess whether a simulated planet would be detected, we fit each simulated set of RVs with a one-planet orbital solution, a linear trend, and a flat line.  We compared these model fits using the Bayesian Information Criterion (BIC) \citep{doi:10.1080/01621459.1995.10476572} in order to determine the simplest model that can provide an adequate fit to the data.  If the BIC values for either the one-planet model fit or the linear trend were smaller than the BIC value for the flat line by at least 10, we concluded that the simulated planet would have been detected.  However, if the flat line was preferred or the difference in BIC was less than 10, we counted this as a non-detection.  We repeated this process for each simulated companion injected into each grid box, using our ``detected/not detected" determinations to calculate the completeness over the entire grid. 

Perhaps unsurprisingly, we find that the average sensitivity to companions in systems with super-Earths discovered via the transit method is significantly less than in systems with RV-detected super-Earths.  This likely reflects the substantially greater investment of RV time required to detect a planet with an unknown orbital period and phase, versus the transit case where these two quantities are known precisely in advance.  RV-only detections must also achieve a higher significance in their measurement of the RV semi-amplitude in order to be considered a secure detection \citep[see representative trend system GJ 273;][]{2009A&A...493..645F,2017AJ....153..208B},
whereas for RV follow-up of transiting planets even marginally significant measurements of this quantity still provide useful constraints on the planet density \citep[see representative trend system Kepler 97;][]{2014ApJS..210...20M}.  We show the resulting completeness maps in Figure 5, with systems discovered using the transit method plotted separately from systems discovered using RVs in order to illustrate the different average sensitivities of these two samples.

\begin{figure*}
\begin{tabular}{ c  c  }
\includegraphics[width=0.5\textwidth]{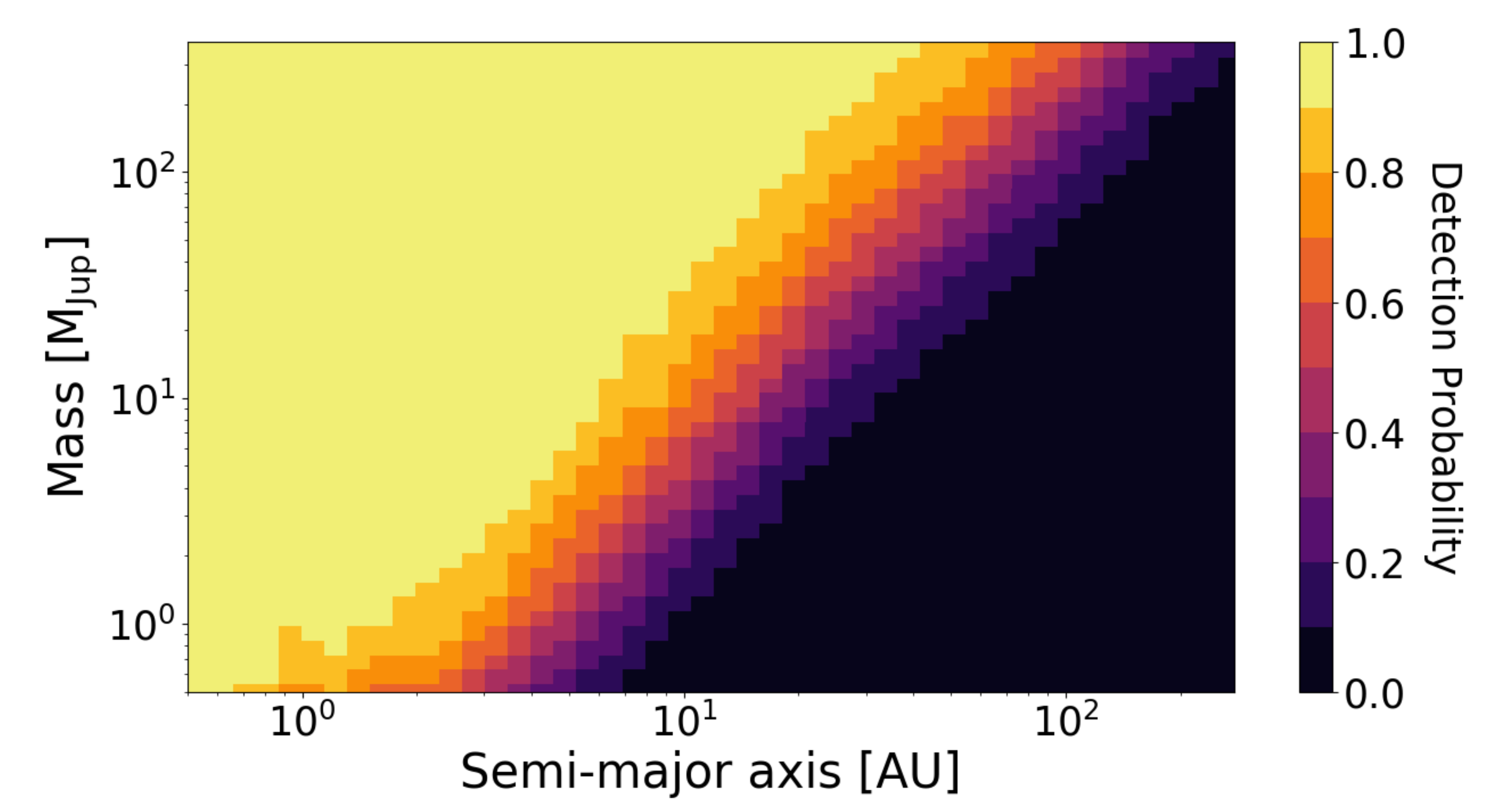}&
\includegraphics[width=0.5\textwidth]{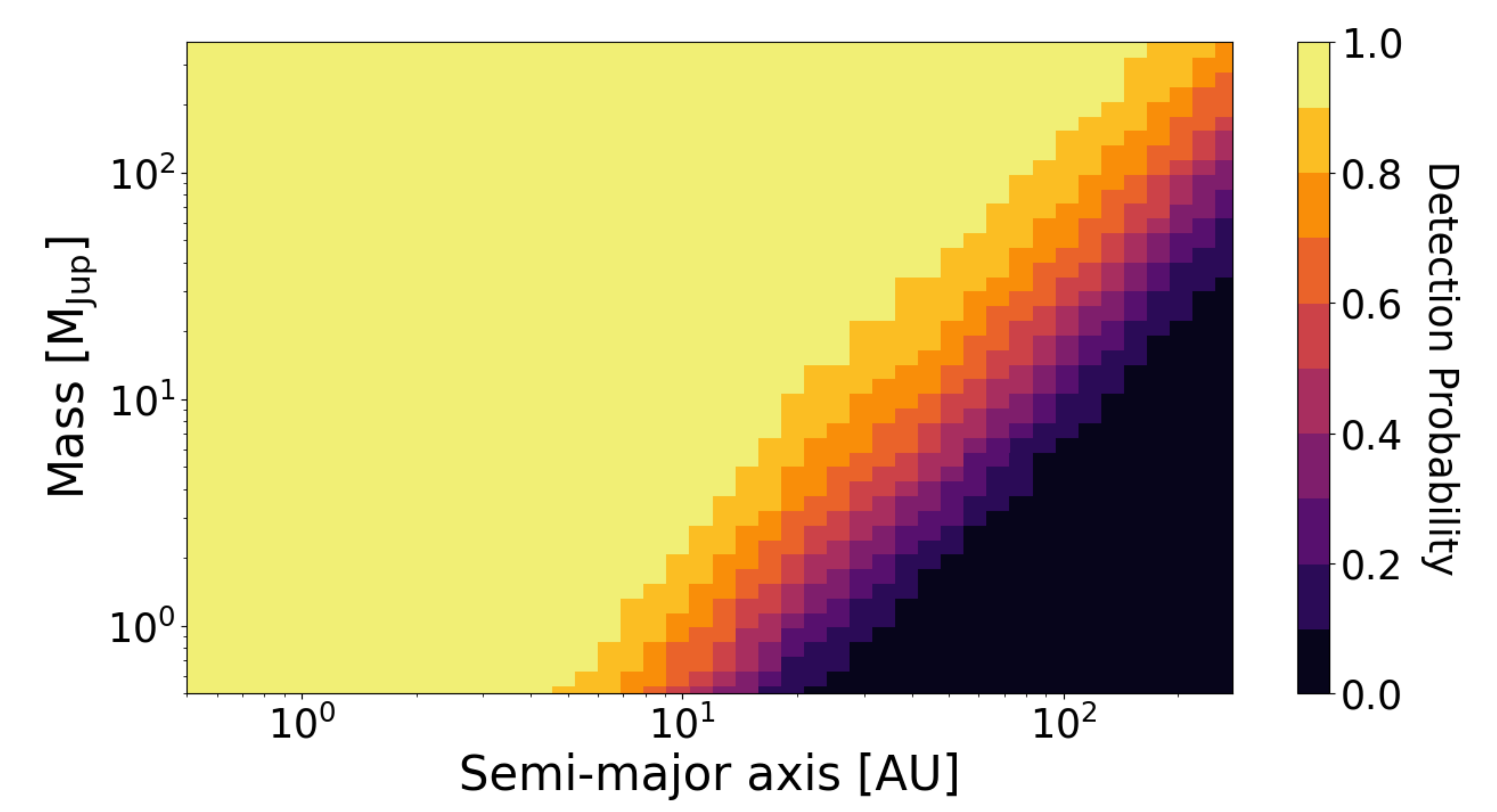}
\end{tabular}
\caption{Sensitivity maps for the systems with super-Earths discovered using the transit method (left) and radial velocity method (right).  Radial velocity detections typically require much more extensive data sets and have longer baselines than observations of transiting planet systems, resulting in different levels of completeness for these two samples.}  
\end{figure*}

\section{Discussion}

\subsection{The Occurrence Rate of Gas Giant Companions}

In this section we utilize our probability distributions (see section 3.4) for each system to determine the underlying distribution and corresponding occurrence rate for the observed population of long period gas giant companions in these systems.  We follow the methodology laid out in \citet{2016ApJ...821...89B}, and present a summary of the steps here.  We first assume that this population of companions is distributed in mass and semi-major axis space according to a double power law of the form $f(m,a) = Cm^{\alpha}a^{\beta}$ \citep[e.g.][]{2002MNRAS.335..151T,2008PASP..120..531C}.  The likelihood for a set of N exoplanet systems is given by:

\begin{equation}
\mathscr{L} =  \Pi^{N}_{i = 1}p(d_i | C, \alpha, \beta)
\end{equation}

\noindent where $p(d_i | C, \alpha, \beta)$ is the probability of the RV dataset given power law coefficients $C$, $\alpha$, and $\beta$.  Assuming that each system has at most one outer companion, this likelihood is then the sum of the probability that a given system contains one planet and the probability that the system contains zero planets.  The probability of a system containing zero planets is given by:

\begin{equation}
p(d_i, 0 | C, \alpha, \beta) = p(d_i | 0)[1 - Z]
\end{equation}

\noindent where $Z$ is the probability that the system contains a planet within a range of masses and semi-major axes (determined by integrating the power law distribution over the specified range), and $ p(d_i | 0)$ is the probability of obtaining the RV dataset given that there is no planet in the system.  

The probability of a system having one companion given their distribution in mass and semi-major axis space is:

\begin{align}
p(d_i, 1 | C, \alpha, \beta) = \int_{a_1}^{a_2} d\ln a \int_{m_1}^{m_2} d\ln m\nonumber \\
 \times \hspace{0.1cm} p(d_i | a, m) \hspace{0.1cm} C m^{\alpha} a^{\beta}
\end{align}

\noindent where $p(d_i | a, m)$ is the probability of a companion being located at a given mass and semi-major axis, which we know from our previously determined probability distributions (see section 3.4).  To determine the likelihood of a given set of $C$, $\alpha$, and $\beta$ given our RV datasets, we combine the probabilities of a system having one planet and a system having zero planets as follows:

\begin{equation}
\mathscr{L} =  \Pi^{N}_{i = 1} \bigg[p_i(d_i, 0 | C, \alpha, \beta) + p_i(d_i, 1 | C, \alpha, \beta)\bigg]
\end{equation}

As in \citet{2016ApJ...821...89B}, we incorporate the probability distributions for all systems in this framework, not just the distributions for systems that have statistically significant trends.  This allows us to treat all systems consistently regardless of whether or not they have a statistically significant trend.  Phrased another way, this allows for the possibility of marginal trend detections, rather than assuming a binary classification system in which any star with a less than $3\sigma$ trend is counted as a non-detection.  For the nine systems hosting exterior gas giant ($>$0.5 M$_{\rm Jup}$) companions on resolved orbits outside 1 AU, we replace the probability distributions calculated from the RV trends with ones where the probability density is concentrated in a single grid point closest to the best fit mass and orbital separation of the resolved companion (see Table 3 for these values).

Two of these systems (GJ 676 and Kepler-454) have both statistically significant trends and resolved gas giant companions, while HD 181433 has two resolved gas giant planet companions.  In these cases, we select the innermost gas giant planet for inclusion in our power-law fit.  We also note that our choice to use the inner vs outer companion in the system with more than one gas giant companion does not impact our derived power law coefficients or occurrence rates.  

We determine the values of $C$, $\alpha$, and $\beta$ that maximize the value of $\mathscr{L}$ by first performing a grid search where we vary each of these power law coefficients, and then carry out a MCMC fit initialized near the location of the optimal grid point.  Because these parameters are both poorly constrained and highly correlated, we find that the use of a preliminary grid search allows us to reliably identify the global maximum and reduces the convergence time in our MCMC chains.  

We next use the results of these power-law fits to calculate an integrated occurrence rate for the observed population of gas giant companions over a range of masses and semi-major axes.  We first calculate the integrated companion frequency separately for systems discovered using the transit method versus the radial velocity method.  Given the significant differences in completeness for these two samples of systems, this allows us to evaluate the degree to which these sensitivities impact the integrated occurrence rates.  We ran the grid search and MCMC analysis of each sample separately.  When we calculated the occurrence rates for these two samples of systems over a mass range of 0.5 - 20 M$_{\textnormal{Jup}}$ and a semi-major axis range of 1 - 20 AU, we found that the occurrence rate of companions in the transiting planet sample is 41$^{+10}_{-10}\%$, while the occurrence rate of the RV planet sample is 34$^{+11}_{-10}\%$. These two values are consistent at the 0.5$\sigma$ level.  We note that the uncertainties on these occurrence rates are dominated by the number of systems in each sample, which are similar (34 for the transiting planet sample, 25 for the RV sample). 

We next calculate the frequency of companions for the combined sample over different ranges in mass and semi-major axis in order to assess how occurrence rates depend on our chosen integration ranges.  Table 6 shows the resulting occurrence rates for the combined sample.  We note that, as in \citet{2016ApJ...821...89B}, the values for the power law coefficients $\alpha$ and $\beta$ vary significantly depending on our chosen integration range as a result of the poorly constrained companion masses and separations in these systems.  However, we find that we obtain consistent results for the integrated occurrence rate for these companions across a wide range of integration ranges.  This is because the strongest constraint we obtain from these data is the total number of companions in these systems, while their locations are poorly constrained. As a result, we find that the preferred values for $C$, $\alpha$, and $\beta$ in our fits are correlated in a way that preserves the total number of companions regardless of the fitting range used.  This stands in contrast to studies examining populations of planets with tightly constrained masses and orbital semi-major axes \citep[e.g.][]{2008PASP..120..531C,2010ApJ...709..396B}, where the values of $\alpha$ and $\beta$ are much better constrained by the data.  For these systems, we would expect the integrated occurrence rate to rise as we increase the range in mass and semi-major axis, reflecting our much better knowledge of the planet occurrence rate density.  This is an important point to consider when comparing our occurrence rate to those from surveys focusing on planets with fully resolved orbits, as we will discuss below. 

\begin{deluxetable}{lccc}
\tabletypesize{\scriptsize}
\tablecaption{Total Occurrence Rates for Companions}
\tablewidth{0pt}
\tablehead{
\colhead{} & \colhead{1 - 10 AU} & \colhead{1 - 20 AU}& \colhead{1 - 50 AU}
}
\startdata
0.5 - 20 M$_{\textnormal{Jup}}$ & 38$\pm$7$\%$ & 39$\pm$7$\%$& 41$^{+8}_{-7}\%$\\
0.5 - 13 M$_{\textnormal{Jup}}$ & 36$^{+7}_{-6}\%$& 41$^{+8}_{-7}\%$& 40$^{+8}_{-7}\%$\\
1 - 20 M$_{\textnormal{Jup}}$ &35$\pm$7$\%$ & 35$\pm$7$\%$ & 38$^{+8}_{-7}\%$\\
1 - 13 M$_{\textnormal{Jup}}$ & 34$\pm$7$\%$& 38$\pm$7$\%$& 39$^{+8}_{-7}\%$ 

\enddata
\end{deluxetable}

Our estimated occurrence rate has a relatively weak dependence on the assumed eccentricity distribution.  As discussed in section 3.4, we draw eccentricities from a beta distribution to calculate the probability distributions for each companion. When we recalculated our probability distributions using either circular orbits or a uniform distribution in eccentricity, we found occurrence rates of 36$\pm$7$\%$ and 40$\pm$7$\%$ respectively. Both of these values are consistent with our nominal  occurrence rate of 39$\pm$7$\%$ calculated using the beta distribution. We also test our assumptions regarding the inclination distribution of the companions.  Previous dynamical studies have shown that for some multi-transiting systems, an exterior companion must have a similar inclination (i.e. Becker $\&$ Adams 2017).  We test an extreme version of this scenario by recalculating all probability distributions for multi-transiting systems using a restricted inclination range of 5 degrees (87.5-92.5 degrees).  We rerun the occurrence rate calculation for all systems using these restricted inclination distributions for the multi-transiting systems, and obtain an occurrence rate of 38$\pm$7$\%$.  We conclude that restricting the inclination for these systems does not significantly impact the occurrence rate.  Finally, we explore how our calculated occurrence rate might be impacted if some of the less significant trends are caused by stellar activity.  We test an extreme scenario in which we assume all systems with $<1\sigma$ trends (15 systems) are caused by stellar activity, by replacing these trends with flat lines.  We found that when we recalculated the occurrence rate over the range 0.5 - 20 M$_{\rm Jup}$ and 1 - 20 AU, it was 37$\pm$7$\%$, consistent with the original occurrence rate of 39$\pm$7$\%$.  We take this a step further and replace all systems with $<2\sigma$ trends (40 systems) with flat lines.  When we rerun the occurrence rate calculation, we find an occurrence rate of 30 (+7 -6)$\%$ over the range 0.5 - 20 M$_{\rm Jup}$ and 1 - 20 AU.  Thus the bulk of the occurrence rate is being driven by stronger trends and resolved companions, as expected, and we conclude that it is unlikely that stellar activity is the cause of a majority of these weaker trends.

\subsection{Comparison to Field Jupiter Analog Occurrence Rates}

We now aim to determine whether the rate of gas giant companions in super-Earth systems is higher or lower than the average occurrence rate for sun-like field stars.  If there is no correlation (positive or negative) between the presence of an inner super-Earth and an outer gas giant companion, we would expect these two rates to be consistent with each other.  

In this study we examine 65 super-Earth-hosting stars, where we define a super-Earth as a planet having either a mass between 1 -- 10 M$_{\Earth}$ or a radius between 1 -- 4 R$_{\Earth}$ depending on the detection method.  Similarly, we define a long period gas giant as a planet with a mass between 0.5 -- 20 M$_{\rm Jup}$ and semi-major axis between 1 -- 20 AU.  When comparing to previous work we vary this definition of long-period gas giant planet to better match individual surveys as discussed below.

There have been several studies that have sought to quantify the frequency of long period gas giant planets, including \citet{2016ApJ...819...28W}, \citet{2016ApJ...817..104R}, and \citet{2016AJ....152..206F}.  \citet{2016ApJ...819...28W} calculate the occurrence rate of Jupiter analogs over the range 0.3 - 13 M$_{\textnormal{Jup}}$ and between 3 - 7 AU for a sample of 202 stars observed as a part of the 17-year Anglo-Australian Planet Search.  For their sample of targets, they only consider planets with fully resolved orbits and find an integrated occurrence rate of 6.2$^{+2.8}_{-1.6}\%$ over this range assuming binomial statistics (i.e., they do not fit a power law distribution).  Integrating our sample over this same mass and semi-major axis range, we find an occurrence rate of 34$\pm7\%$, which differs from the \citet{2016ApJ...819...28W} value by 3.7$\sigma$ (Figure 6).  We note that for this occurrence rate calculation, in a system with two gas giant companions (either both resolved or one resolved, one statistically significant trend), if the innermost resolved companion does not fall within this integration range, we use the outer companion (this was the case for systems HD 181433, GJ 676, and Kepler-454).  As stated previously, the inclusion of either the closer resolved gas giant companions or the farther out signals does not affect our occurrence rate calculations over the range 0.5 - 20 M$_{\rm Jup}$ and 1 - 20 AU.

In order to determine whether or not the difference between our occurrence rate of 34$\pm$7$\%$ and the 6.2$^{+2.8}_{-1.6}\%$ rate is meaningful, we must consider the possible biases introduced by our decision to consider trends rather than limit our study to companions with fully resolved orbits.  Specifically, our occurrence rate is partially derived from a population of planets with probability distributions extending over a wide range of masses and semi-major axes. This means that when we integrate over the relatively narrow range used in the Wittenmyer study, our occurrence rate may be inflated by the inclusion of planets whose probability distributions overlap with this integration range, even though the planets themselves are in fact located on more distant orbits. 

\begin{figure}[h]
\centering
\includegraphics[width=0.5\textwidth]{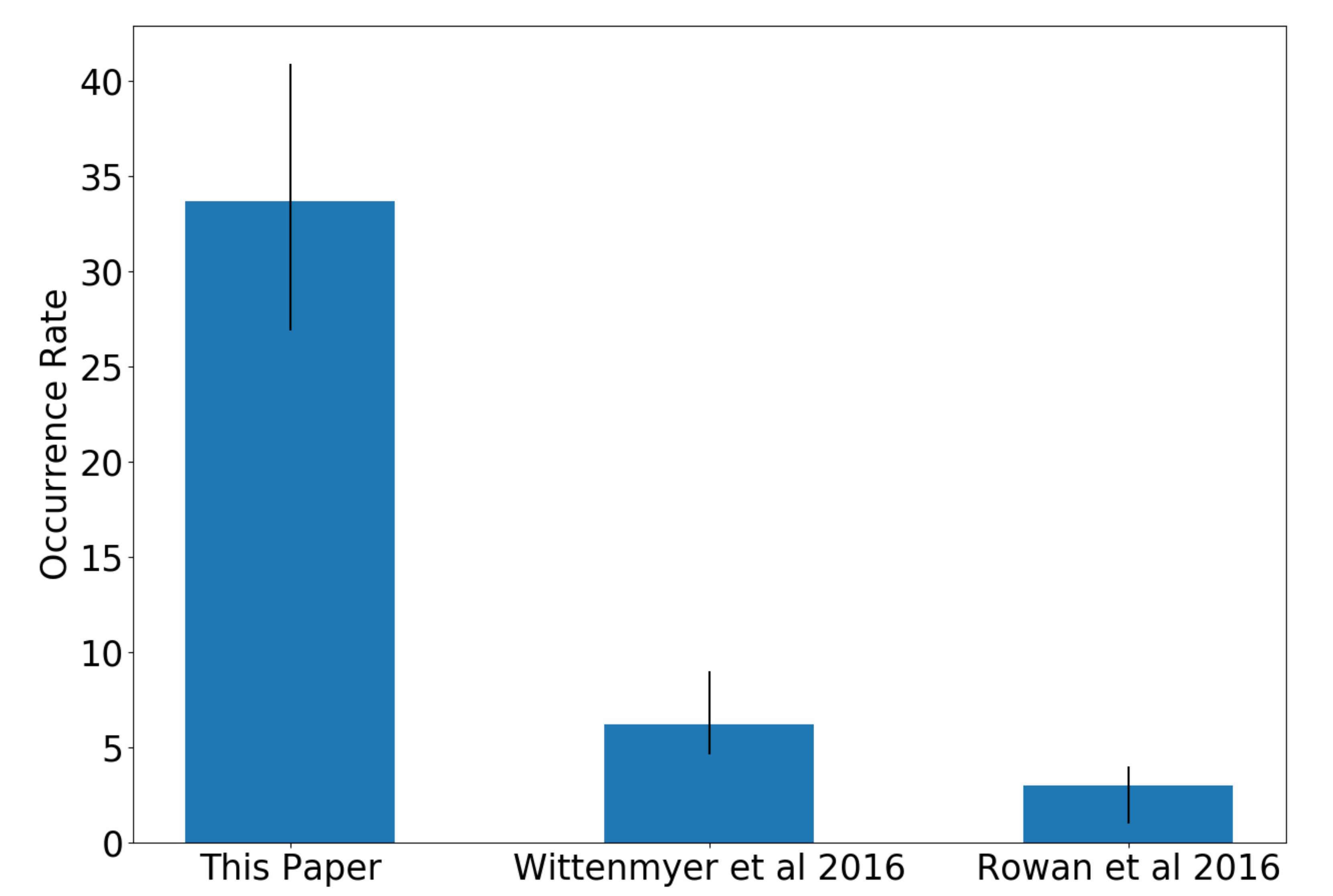}
\caption{Compared to the Jupiter analog occurrence rate estimates published in \citet{2016ApJ...819...28W} and \citet{2016ApJ...817..104R}, this study finds a higher occurrence rate of distant gas giant planets in super-Earth systems than would be expected just based on chance. Occurrence rate integration ranges are 0.3 - 13 M$_{\rm Jup}$ and 3 - 7 AU for this paper and \citet{2016ApJ...819...28W}, and 0.3 - 3 M$_{\rm Jup}$ and 3 - 6 AU for \citet{2016ApJ...817..104R}.}
\end{figure}

To quantify this effect, we assume that planets in our sample are distributed according to a negative power law in mass, and a flat power law in semi-major axis.  Specifically, we adopt the $\alpha$ value of -0.31 from \citet{2008PASP..120..531C}, and a $\beta$ value of 0. We choose not to adopt the Cumming et al power law coefficient for semi-major axis ($\beta$ = 0.26), as this coefficient was derived from a fit to the population of gas giant planets inside 3 AU.  This fit indicates that the frequency of these planets rises with increasing semi-major axis, but this is inconsistent with current constraints from both RV and direct imaging surveys \citep{2016ApJ...821...89B,2016PASP..128j2001B,2016ApJ...819..125C,2018arXiv180210132B}, which prefer much flatter distributions at large semi-major axes. In \citet{2016ApJ...821...89B} we found that for the population of gas giant planets with long-period companions, the occurrence rate of these companions decreases with increasing semi-major axis.  While the current small sample size of directly imaged planets makes it difficult to determine their mass and semi-major axis distribution, their overall low occurrence rate indicates that a rising power law in semi-major axis is likely not applicable at wide separations.  

For each of the nine systems with a statistically significant trend that do not have a resolved companion in this mass and semi-major axis range, we draw from the $\alpha$ = -0.31, $\beta$ = 0 power law distribution until we have generated a sample of 100 simulated planets with a cutoff mass of 20 M$_{\textnormal{Jup}}$ that lie within the favored region of mass/semi-major axis parameter space where the probability of there existing a planet given the RV trend is greater than the probability of there being no planet given the RV trend.  For each system we then count the fraction of planets that fall within the range $3-7$ AU and $0.3-13$ M$_{\textnormal{Jup}}$.  For the resolved companions, we count three companions that fall within this range and were thus included in the occurrence rate calculation.  Averaging across all of the trend and resolved companion systems, we find that 61\% of our simulated planet population lies inside this range.  If we rescale our occurrence rate to account for the fact that $\sim$2/5 of our occurrence rate might be attributed to companions outside the 3 - 7 AU semi-major axis range, we would then derive a corrected occurrence rate of 21$\%\pm4\%$ for our sample over this semi-major axis range.  This reduced occurrence rate is still inconsistent with the Wittenmyer et al. value at the 3$\sigma$ level. 

We note that \citet{2016ApJ...819...28W} mentions that they find 45 trends in their sample of systems.  However, these trends could not be Jupiter analogs located between 3 - 7 AU because the eight year minimum baseline (the specified cutoff for system inclusion in the 202 star sample) would have been sufficient to determine whether or not the orbit was within 3 - 7 AU (Wittenmyer 2018, email commun.).  Furthermore, while eight years was the nominal cut off, over 90$\%$ of systems had baselines above 4000 days ($\sim$11 years), while a majority of the sample had baselines over 6000 days ($\sim$16.5 years).  We therefore conclude that Wittenmyer et al.'s decision  to ignore trends in their occurrence rate calculation is unlikely to result in systematic differences in sensitivity as compared to our study. 

We next consider results from other studies that provide independent estimates of the frequency of long period gas giant planets around nearby stars.  \citet{2016ApJ...817..104R} estimate the occurrence rate of Jupiter analogs using a sample of 1122 stars, where they define a Jupiter analog as a planet with a mass between $0.3-3$ M$_{\textnormal{Jup}}$ and semi-major axis between $3-6$ AU.  As with Wittenmyer et al., they only consider planet detections with fully resolved orbits in their analysis.  Over this range, they find an occurrence rate of $1-4\%$.  While this integration range has a relatively strict mass limit, we note that previous power law fits to the population of RV-detected planets consistently indicate that lower mass gas giants are more common than higher mass gas giants \citep{2008PASP..120..531C,2016ApJ...821...89B}, suggesting that their upper bound of 3 M$_{\textnormal{Jup}}$ versus our upper bound of 13 M$_{\textnormal{Jup}}$ is unlikely to change this integrated occurrence rate very much.  We therefore conclude that our occurrence rate is likely higher than the rate from this study as well, with the same caveats as for the Wittenmyer et al. comparison.  

For our last comparison we turn to \citet{2016AJ....152..206F}, who calculated the frequency of long period planets between $1.5-9$ AU and $0.01-20$ M$_{\textnormal{Jup}}$ using transit detections from the Kepler photometry.  Unlike the previous two radial velocity studies, a majority of the long period planets in their sample have just one observed transit.  Although this study is able to place some loose constraints on the orbital periods of these planets based on their measured transit durations, these constraints are nearly as broad as those for our radial velocity trend systems.  For this parameter space \citet{2016AJ....152..206F} find an occurrence rate density of 0.068$\pm$0.019, corresponding to an integrated occurrence rate of $92.5 \pm 25.7\%$.  Over a similar semi-major axis range and a more limited mass range ($1-10$ AU and $0.5-20$ M$_{\textnormal{Jup}}$), we find an occurrence rate density of 0.045 $\pm$ 0.009 and an integrated occurrence rate of 38$\pm7\%$.  While these two occurrence rate densities are formally consistent, three-quarters of \citet{2016AJ....152..206F}'s sample consists of planets whose estimated masses are less than 0.2 M$_{\textnormal{Jup}}$, whereas all of our candidate companions have minimum masses higher than this threshold.  We therefore conclude that there is relatively little overlap between the two planet samples, making this comparison less relevant than the studies by \citet{2016ApJ...819...28W} and \citet{2016ApJ...817..104R}.

\subsection{Frequency of Super-Earths in Jupiter Analog Systems}

Now that we have determined the frequency of long period gas giants in super-Earth systems, we can ask what fraction of the long period gas giants orbiting field stars are drawn from this population (i.e., what fraction of long period gas giants have inner super-Earths?).  We can express this as a conditional probability: 

\begin{equation}
p(SE | LPG) =  \frac{p(LPG | SE) \times p(SE)}{p(LPG)}
\end{equation}
where $p(SE)$ is the probability that a given star hosts a super-Earth and $p(LPG)$ is the probability of hosting a long period gas giant planet. We note that \citet{Zhu18} present a similar equation and provide an independent estimate of this probability, which we discuss in more detail below.  If we take our estimate of the frequency of Jupiter analogs in super-Earth systems, we can use this equation to calculate what fraction of Jupiter analogs host inner super-Earths.  For consistency we integrate the conditional occurrence rate P(LPG$|$SE) from our study over the same mass and semi-major axis range as \citet{2016ApJ...819...28W}, resulting in P(LPG$|$SE) = 34$\pm$7$\%$ and P(LPG) = 6.2$^{+2.8}_{-1.6}\%$.  We take the super-Earth occurrence rate of P(SE) = 30$\pm$3$\%$ from \citet{Zhu20182}.  Using Equation 7 we find that the conditional probability that there is at least one super-Earth in a system hosting a Jupiter analogs is P(SE$|$LPG) = 164$^{+57}_{-83}\%$. Although our definition for this quantity disallows values greater than 100$\%$, it is likely that some of the trends that we include in our Jupiter analog occurrence rate calculation may be caused by planets beyond 7 AU.  As discussed in Section 4.2, accounting for this correction can reduce our Jupiter analog occurrence to 21$\pm$4$\%$ if we assume a flat power law distribution in semi-major axis.  Using this lower value in the conditional probability calculation, we find P(SE$|$LPG) = 102$^{+34}_{-51}\%$.  Given the relatively large uncertainties on both the occurrence rates used in Equation 7 and the estimated semi-major axis distribution of planets at these large separations, it is difficult to obtain a precise constraint on the occurrence of super-Earths in Jupiter analog systems P(SE$|$LPG). Nevertheless, this result does appear to suggest that most if not all systems hosting a Jupiter analog also host one or more inner super-Earths.

\citet{Zhu18} recently published an independent estimate of the occurrence rate of gas giant companions in super-Earth systems P(LPG$|$SE) as well as the conditional occurrence rate of super-Earths in Jupiter analog systems P(SE$|$LPG).  They also find that super-Earth systems host more cold Jupiters than would be expected based on chance alone, and conclude that most if not all systems hosting cold Jupiters also host inner super-Earths.  \citet{Zhu18} estimate P(LPG$|$SE) using a sample of 22 transiting super-Earth systems from Kepler with radii between 1 -- 4 R$_{\Earth}$ and 32 RV systems with masses $<$20 M$_{\Earth}$.  Assessing the overlap between our two samples, we include 21 of their 22 transiting planet systems in our study and 10 of their RV systems.  While we missed Kepler-89 in our study, for our RV sample, only 10 of the RV systems in \citet{Zhu18} meet our requirement that $m$sin$i$ be less than 10 M$_{\rm \Earth}$.  For the purpose of this calculation, \citet{Zhu18} define cold Jupiters as planets with $m$sin$i$ $>$ 0.3 M$_{\rm Jup}$ and orbital periods $>$ 1 year, and count the number of cold Jupiters in these systems as reported in the NASA Exoplanet Archive.  This results in an occurrence rate of P(LPG$|$SE)=32$\pm$8$\%$ assuming 100$\%$ completeness.  Using an equivalent conditional probability to our Equation 7, \citet{Zhu18} set P(LPG) = 10$\%$ and P(SE) = 30$\%$ and conclude that P(SE$|$LPG) = 90$\pm$20$\%$.

It is reassuring that Zhu $\&$ Wu's estimated values for P(LPG$|$SE) and P(SE$|$LPG) are consistent with our values despite differences in both sample selection and methodology.  However, we note that more quantitative comparisons are difficult given that the authors of this study do not define an explicit upper boundary in mass and semi-major axis for their cold Jupiter sample.  Our results demonstrate that the sensitivity of these RV data sets to cold Jupiters can vary greatly from system to system as a result of differences in baseline, cadence, and sample size.  Without an upper boundary in semi-major axis, it is difficult to evaluate whether or not Zhu \& Wu's implicit assumption of 100\% completeness is valid for their chosen sample.  Similarly, if we wish to compare our value for P(LPG$|$SE), which requires integration over a finite range in mass and semi-major axis space, to that of Zhu \& Wu it is unclear what limits of integration we should use.  

As noted above, our study also uses a different mass cutoff for our RV super-Earth sample selection than that utilized by Zhu $\&$ Wu.  While both studies use the same radius range for the transiting planet sample, we require our RV super-Earths to have  $m$sin$i$ between $1-10$ M$_{\Earth}$ while \citet{Zhu18} use a more generous range of $1-20$ M$_{\Earth}$.  As a result, 40$\%$ of their 54 super-Earths systems contain planets with $m$sin$i>$ 10 M$_{\Earth}$.  This alternate definition increases the likelihood that their sample contains a substantial fraction of Neptune-mass planets (see section 2 for a discussion of this probability for our RV sample).  However, while this is an important difference, the fact that we obtain consistent estimated values for P(LPG|SE) in both studies suggests that this choice may not substantially alter their conclusions.

\subsection{Implications of Our Results for Super-Earth Formation and Migration Models}

Although it is difficult to make quantitative comparisons without a better understanding of the power law distribution for the long period gas giant planets in our sample, our results indicate that there is a higher occurrence rate for gas giants in systems hosting inner super-Earths than for field stars. Furthermore, if we take our integrated occurrence rate of 34$\pm7\%$ between $3-7$ AU and $0.3-13$ M$_{\textnormal{Jup}}$ at face value, as well as the overall occurrence rates of super-Earths and long-period gas giant planets \citep{2010Sci...330..653H,2013ApJ...766...81F,2013PNAS..11019273P,2016ApJ...819...28W,2016ApJ...817..104R,2018arXiv180209526Z}, Equation 7 would suggest that a significant majority of long period gas giant planets have inner super-Earths. 

\subsubsection{Long Period Gas Giants Do Not Hinder Super-Earth Formation, Did Not Migrate Large Distances} 

The apparent correlation between the occurrence of inner super-Earths and outer gas giants suggests that gas giant companions do not hinder super-Earth formation, either by cutting off the flow of solids to the inner disk, stirring up the velocity distribution of these solids, or by preventing super-Earths formed at larger separations from migrating inward \citep{2011LPI....42.2585W,2015PNAS..112.4214B,2015ApJ...809...94M,2015ApJ...800L..22I}.  We also note that four of the super-Earth systems in our sample contain additional gas giant companions inside 1 AU, which were not included in our estimate of the frequency of exterior long period companions. These four inner gas giants are located immediately adjacent to and in some cases in between the super-Earths in these systems, providing additional support for the idea that the presence of gas giants does not disrupt super-Earth formation. It is likely that these inner gas giants with super-Earth companions did not undergo large scale migration: high eccentricity migration of any kind would destroy the orbits of inner planets \citep[e.g.][]{2015ApJ...808...14M,2016AJ....152..174A}, and disk-driven migration would lead to resonant locking among the giant planets and the super-Earths, resulting in a orbital architecture reminiscent of the Galilean satellites.

If the scenario for the solar system presented in \citet{2015PNAS..112.4214B} is correct, this would also suggest that the long-period gas giant companions in these systems did not undergo large-scale inward migration.  A gas giant undergoing long-range migration \citep[in][Jupiter migrated from 6 AU to 1.5 AU]{2015PNAS..112.4214B} can capture solid materials in resonance, driving both the inward transport and the excitation of eccentricities. The ensuing collisional cascade among the resonant planetesimals shuttles newly formed super-Earths inward onto their host stars.  In this scenario, the total mass participating in this cascade must be large enough to appreciably alter the orbital semi-major axes of inner super-Earths (i.e., it must have a total mass comparable to that of the planets). The amount of solids participating in cascade depends on the distance over which the outer gas giants migrates: $M_{\rm swarm} \sim f_{\rm solid} \times ({\rm nebula\,mass}) \times ({\rm migration\,distance\,/\,size\,of\,the\,disk})$---where we assume a gas disk surface density profile of $\propto 1/r$ with $r$ being the stellocentric distance, and $f_{\rm solid}$ is the solid-to-gas mass ratio. If, for example, the outer gas giant migrates only $\sim$1 AU for $f_{\rm solid} = 1/100$, nebula mass of 0.01$M_\odot$ and disk size of 30 AU, $M_{\rm swarm} \sim 1M_\oplus$. This is not enough mass to shepherd inner super-Earths onto a host star, rendering this mechanism ineffective in systems where giant planets remain relatively close ($\sim <$1 AU) to their formation locations.

\subsubsection{A Single, Long Period Gas Giant Is Unlikely to Destroy Inner Super-Earths}

Most systems we identify with significant evidence for an outer companion (see Tables 3 and 5) have strongly coupled inner planets that are resilient against the perturbation of an outer companion (i.e., their mutual secular precession frequency is higher than the outer perturber's precession frequency). Following the procedure outlined in 
Pu \& Lai (2018), we find that the root-mean-square eccentricities (and mutual inclinations) from the secular perturbation by an outer gas giant is limited to, at maximum, just a few percent over a wide range of perturber eccentricities (0--0.4) and inclinations (0--40$^o$)\footnote{The strength of the secular perturbation is strongly sensitive to the ratio of the orbital distances of the perturber and the inner planet. In our calculation, we adopt the minimum orbital distance of the perturber and its corresponding mass to quantify the maximum possible forced eccentricities and mutual inclinations. We note that the relatively small separation of the outer companion in the CoRoT-24 system can drive the system to orbital instability for perturber eccentricities larger than $\sim$0.4. If we instead sample a range of distances and masses of CoRoT-24 d contained within the 1$\sigma$ probability (see Figure 4), the root-mean-square eccentricities and inclinations drop to just a few percent.}.
The ability of these inner super-Earth systems to remain stable---and roughly coplanar---against the perturbation of the outer perturber, and the robustness of our inferred occurrence rate Jovian companion over a range of eccentricity and 
inclination distributions (see Section 4.1)
suggest that the orbital architecture of inner super-Earths is largely unaffected by the presence of an outer gas giant companion.

One notable exception is Kepler 454, which has an inner super-Earth at $\sim$10 days and a Jovian companion at $\sim$500 days as well as an RV acceleration consistent with an outer giant companion at $\sim$10 AU. While the innermost super-Earth couples more strongly with its closest Jovian neighbor, this Jovian planet couples more strongly with its outer giant companion. Calculating the expected root-mean-square eccentricities from the secular perturbations, we find that an outermost companion with an eccentricity greater than $\sim$0.1 would drive the inner planets to orbital instability. We expect there to be similar limits on the eccentricities of outer gas giants in systems such as GJ 676 A and HD 181433, which harbor multiple gas giants.  In fact, 
\citet{Campanella13} argue that the HD 181433 planets must be in a mean motion resonance in order to ensure orbital stability.
The ability of multiple outer gas giants to dynamically excite inner systems has been studied extensively by 
\citet{Hansen15} for the specific case of 55 Cnc and more generally by 
\citet{Hansen17}. Similar system-by-system stability analyses using full N-body calculations would help to place stronger constraints on the eccentricities and inclinations of the possible outer companions we report here.

\subsubsection{Outer Gas Giants Provide Constraints on Protoplanetary Disk Properties}

We next consider whether or not the presence of an outer gas giant can be used to constrain the properties of the protoplanetary disks from which these systems formed. Large solid mass content is considered one such property, facilitating the growth of grains to planetesimals \citep[e.g.][] {2004ApJ...601.1109Y,2012A&A...539A.148B}, accelerating the growth of cores by pebble accretion \citep[e.g.][]{2012ApJ...747..115O,2014A&A...572A.107L,2018arXiv180600487L}, 
and speeding up the final assembly by giant impact \citep[e.g.][]{2015MNRAS.453.1471D}.\footnote{To be more precise, the growth of planetesimals and cores is governed by the ``local'' concentration of solids; in other words, what matters is the solid-to-gas mass ratio at the site of such growth, not necessarily the bulk mass ratio.} Observationally, both gas giants and super-Earths (here defined as planets with mass of 1--10 $M_\oplus$ and/or with radii of 1--4 $R_\oplus$) are found to occur more frequently around metal-rich stars \citep{2005ApJ...622.1102F,2018AJ....155...89P}. Here we consider whether the metallicity of the host star---used as a proxy of the total solid content in the natal disk---is correlated with the occurrence of gas giant companions to inner super-Earths.

As a test of this question, we divide our sample into systems that have greater than $3\sigma$ trends (systems listed in bold in Table 1, and resolved companions (systems listed in Table 3), which we refer to hereafter as ``Companion Systems'', and those that do not, referred to as ``No Companion Systems'', and compare the error weighted averages of the stellar metallicities between these two samples.  For the transiting planet sample, we find average metallicities of 0.207$\pm$0.005 and 0.076$\pm$0.003 for the companion/no companion systems, a 22$\sigma$ difference.  For the RV sample, the error-weighted average metallicities are 0.346$\pm$0.010 and -0.292$\pm$0.005 for the companion/no companion systems, a $57\sigma$ difference.  We note that in this comparison we do not include M stars GJ 3293 and GJ 163. This is because we wanted to use metallicities for M stars derived using only IR techniques since they are consistent and calibrated on FGK stars, and neither of these M stars had metallicities derived using these techniques.  However, at present the ``companion" and ``no companion" RV sample averages are significantly influenced by a small fraction of systems that have small metallicity uncertainties. In addition, a caveat for this comparison for both the RV and the transit samples is the fact that error weighted averages assume that the scatter in metallicity is simply due to measurement uncertainties as opposed to intrinsic astrophysical scatter, which is not the case here given the typically small metallicity uncertainties.  Furthermore, these comparisons involve small sample sizes, which is not explicitly accounted for in these error weighted averages.     We thus compare these distributions using the Anderson-Darling test, which takes both the intrinsic astrophysical dispersion of the metallicities as well as sample size into account.  Before applying this test we remove all M stars from the sample, which typically had large uncertainties on stellar metallicities in comparison to other stars in the sample, since this test becomes more appropriate the more uniform the sample and the smaller the measurement uncertainties.  After applying the Anderson-Darling test, we find that the probability of the companion/no companion samples coming from the same parent distribution is p-val = 0.0027 and p-val = 0.0286 for the RV and transiting samples, corresponding to a significance of 3.0$\sigma$ and 2.2$\sigma$, respectively.  While the significance of this comparison is substantially lower than that estimated from comparing the error weighted averages, given the potential biases of the error-weighted averages presented above this is a more appropriate assessment of the significance of the differences between the companion/no companion metallicities.  We conclude that these comparisons for the RV and transit samples suggest that super-Earth systems around metal-rich stars are more likely to have outer companions than their metal-poor counterparts.  We plot the distribution of metallicities for companion/no companion systems for each sample in Figure 7, and the metallicities of both RV and transiting samples versus trend significance in Figure 8. We note that the sample of stars in the RV sample typically have lower masses and correspondingly lower metallicities than stars in the transiting planet sample, with an error weighted average metallicity of -0.165$\pm$0.004 and 0.109$\pm$0.003 for the RV and transiting samples respectively.

\begin{figure}[h]
\begin{tabular}{ c   }
\includegraphics[width=0.5\textwidth]{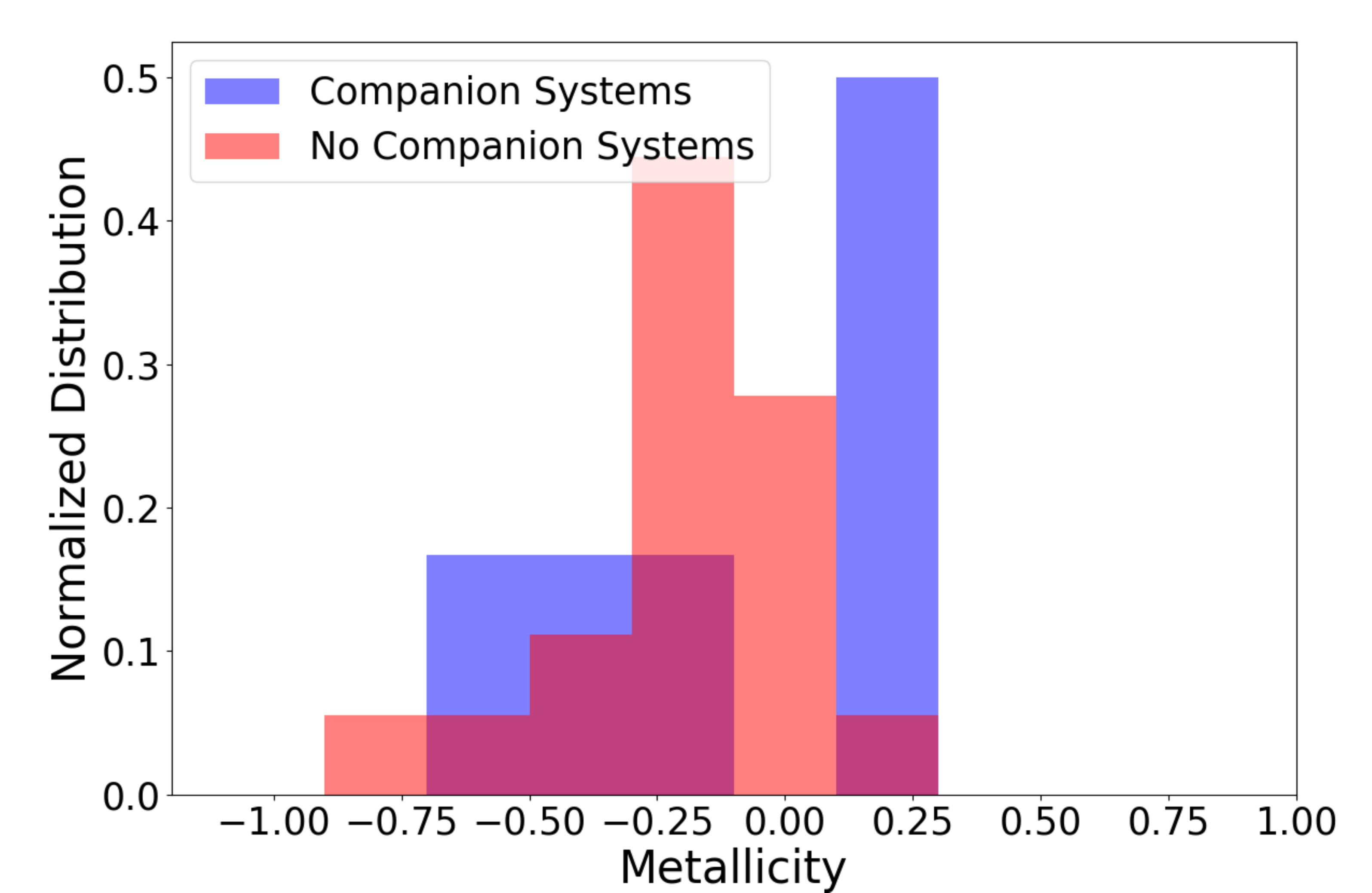} \\
\includegraphics[width=0.5\textwidth]{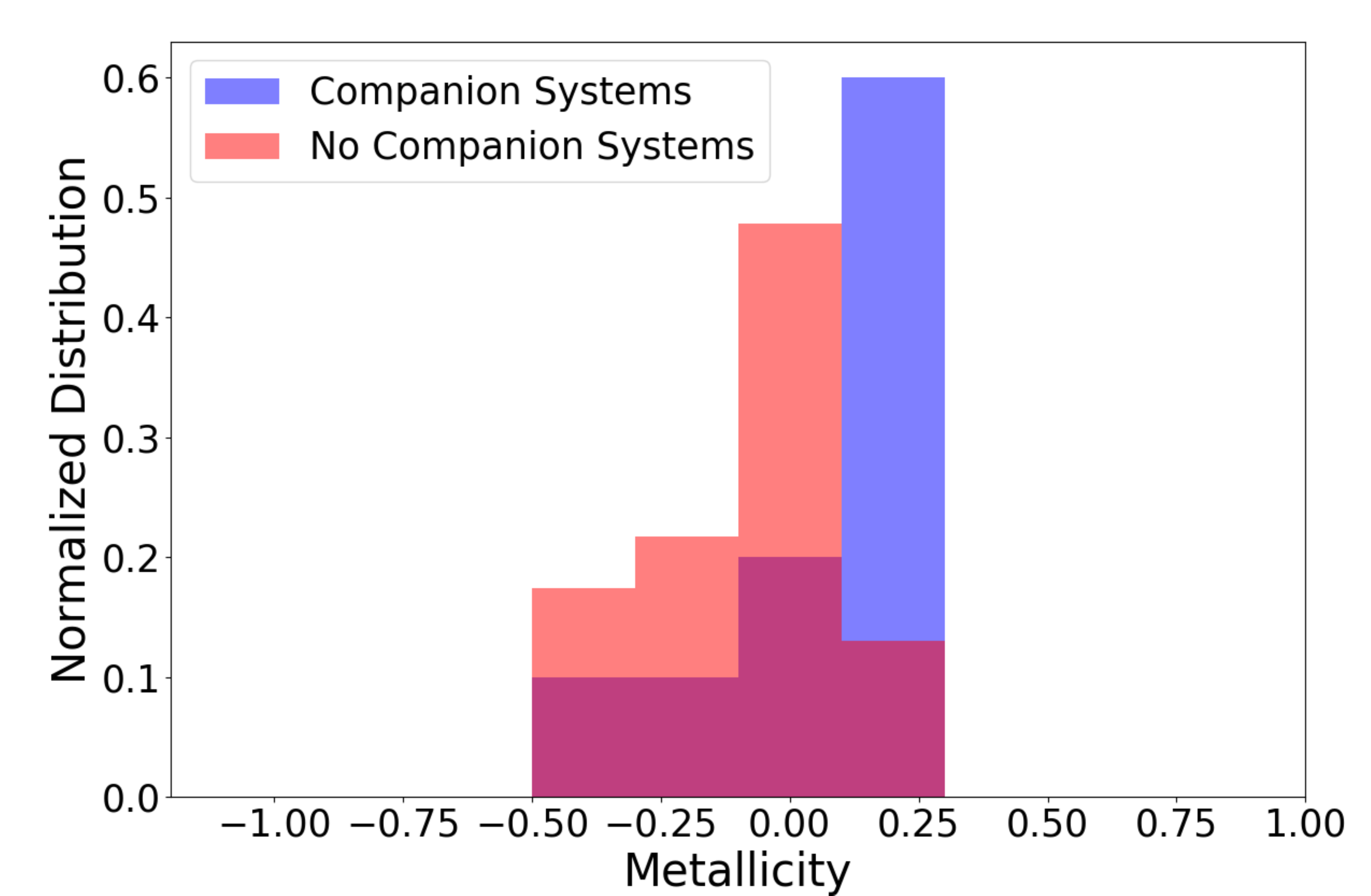} \\
\end{tabular}
\caption{Distributions of stellar metallicities for systems with and without $>3\sigma$ trends and resolved companions.  Top:  RV only sample, with 6 systems in the ``Companion'' sample and 19 systems in the ``No Companion'' sample.  Bottom:  Transit only sample with 10 systems in the ``Companion''sample and 24 systems in the ``No Companion'' sample. }
\end{figure}

\begin{figure}[h]
\includegraphics[width=0.48\textwidth]{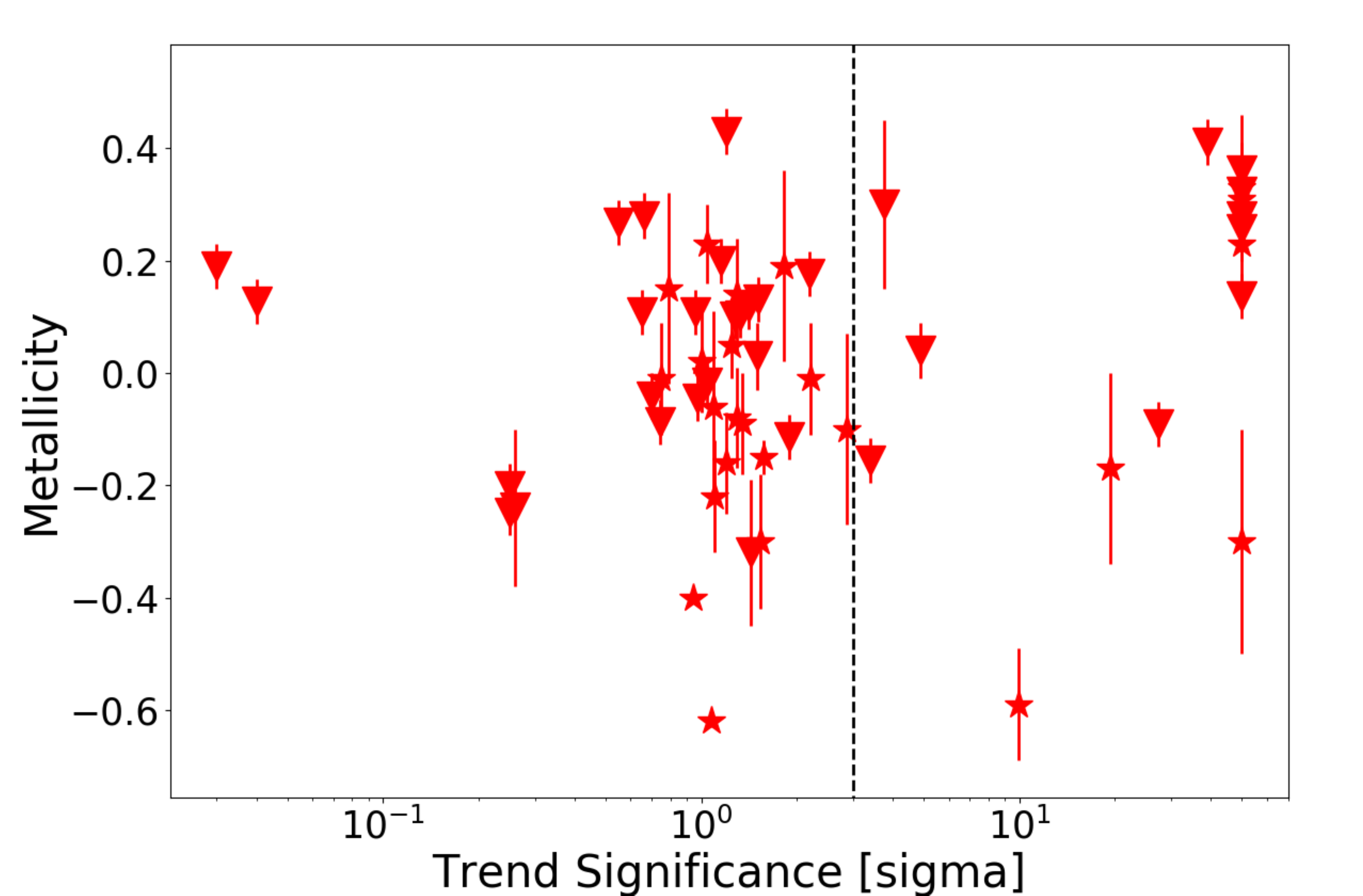} 
\caption{Stellar metallicities plus uncertainties versus trend significance for the RV sample (star symbols) and the transiting sample (triangle symbols).  The black dashed line indicates a 3$\sigma$ trend.  We assign all resolved companions a significance of 50$\sigma$ in order to include them on this plot.}
\end{figure}

We next consider whether or not there is any correlation between the presence of an outer gas giant companion and the mass of the host star.  Observations of young stars indicate that disk mass appears to be correlated with stellar mass, albeit with large intrinsic variance \citep{2013ApJ...771..129A,2016ApJ...831..125P}.  As with stellar metallicity, we expect that disks with higher overall masses will have a correspondingly higher surface density of solids.  However, the benefits of this higher surface density for giant planet formation may be partially negated by the shorter average lifetimes of disks around more massive stars \citep{2015A&A...576A..52R}.  This might affect our estimates of the companion frequency in RV versus transiting planet systems, as these two samples have different stellar mass distributions.  We find that while only one star in the transiting planet sample is an M dwarf (LHS 1140) out of 34 systems, ten targets are M dwarfs in the RV sample (out of 25).  We calculate the occurrence rates for the combined RV and transiting planet sample without the M stars and with M stars only over an integration range of 0.5 -- 20 M$_{\rm Jup}$ and 1 -- 20 AU, and find occurrence rates of 37$\pm$8$\%$ and 44$\pm$17$\%$ respectively.  These occurrence rates are consistent with the occurrence rate of the total sample of 39$\pm$7$\%$ to $<0.1\sigma$ and 0.3$\sigma$ respectively.

\section{Conclusions}
We collected published RV data for a sample of 65 systems hosting at least one inner super-Earth planet in order to search for massive, long-period companions.  We detect these distant companions as long term trends in the RV data when the orbital period of the companion is shorter than the system RV baseline.  Out of our sample of 65 systems, we found 14 systems that had statistically significant trends.  Two of these systems had resolved stellar companions that could potentially have caused the observed trends, while three more systems had trends that were likely due to stellar activity.  We removed these five systems from subsequent analysis, leaving nine systems with statistically significant trends indicating the presence of an outer companion.  Three of these trends are identified here for the first time, while six were previously reported in the literature. We also identify 10 previously published resolved gas giant companions ($>$ 0.5 M$_{\rm Jup}$ and 1 -- 20 AU) in our sample of systems.  We report two new candidate planets in systems HD 156668 and HD 175607, but do not include these in our statistical study as they lie below our mass cutoff with minimum masses of 31 M$_{\Earth}$ and 24 M$_{\Earth}$ respectively. We also recover a fully resolved periodic signal in HD 1461 that appears to be caused by stellar activity, as reported in \citet{2016ApJ...821...89B}. 

We compute 2D probability distributions in mass and semi-major axis space for each system in our sample with a radial velocity trend, where we use the duration and shape of the trends to place lower limits on allowed ranges of mass and separation.  We use a combination of new and archival AO imaging at infrared wavelengths to place a corresponding upper limit on the allowed masses and separations of these companions.  We find that the error-weighted average stellar metallicities of systems with gas giant companions are 5.9$\sigma$ higher than those without companions for our transiting planet sample, and 6.5$\sigma$ higher for our RV sample, in good agreement with the well-established metallicity correlation from RV surveys of field stars.  

We fit the observed companion distributions with a double power law in mass and semi-major axis, and integrate this power law over 0.5 -- 20 M$_{\textnormal{Jup}}$ and 1 -- 20 AU to find an occurrence rate of $39\pm7\%$.  We then compare our occurrence rate for these companions to similar occurrence rates for long period gas giant planets from radial velocity surveys of sun-like field stars. We find that super-Earth systems appear to have more gas giant companions than we would expect to see by chance alone, even after accounting for the additional uncertainty introduced by our inability to pinpoint the precise locations of these companions for systems with radial velocity trends.  The high occurrence rate of long period ($>$1 AU) gas giants in super-Earth systems in turn implies that a significant majority of the long period gas giants identified in radial velocity surveys of field stars likely host inner super-Earths.  We therefore conclude that the presence of an outer gas giants does not hinder super-Earth formation, as proposed in some previous theoretical studies.  To the contrary, our data suggest that these companions may either actively facilitate super-Earth formation or simply serve as a fossil record of early disk conditions that were particularly favorable for planet formation over a wide range of semi-major axes.

\section*{}
This work was supported by NSF CAREER grant 1555095, and was based in part on observations made at the W.M. Keck Observatory. We extend special thanks to those of Hawaiian ancestry on whose sacred mountain of Mauna Kea we are privileged to be guests. EJL is supported by the Sherman Fairchild Fellowship from Caltech. 

\bibliography{super_earth_trends_arxiv_finalv}{}

\end{document}